\documentclass[preprint,12pt]{elsarticle}

\usepackage{amssymb}
\usepackage{hyperref}
\usepackage{subfigure}
\usepackage{graphicx}
\usepackage{amsmath}
\journal{Nuclear Physics A}

\begin{document}

\begin{frontmatter}



\title{Two-Photon, Two-gluon and Radiative Decays of Heavy Flavoured Mesons}


\author{Arpit Parmar$^*$\footnote{arpitspu@yahoo.co.in}, Bhavin Patel$^\dag$\footnote{azadpatel2003@yahoo.co.in} and P C Vinodkumar$^*$\footnote{pothodivinod@yahoo.com}}

\address{$^*$Department of Physics, Sardar Patel University, Vallabh Vidyanagar-388120, Gujarat, INDIA.\\$^\dag$LDRP Institute of Technology and Research, Gandhinagar-382015, Gujarat, INDIA.}

\begin{abstract}
Here we present the two-photon and two-gluon decay widths of the S-wave ($\eta_{Q\in c,b}$) and P-wave ($\chi_{Q\in c,bJ}$) charmonium and bottonium states and the radiative transition decay widths of $c\bar c$, $b\bar b$ and $c\bar b$ systems based on  Coulomb plus power form of the inter-quark potential ($CPP_\nu$) with exponent $\nu$. The Schr$\ddot{o}$dinger equation is solved numerically for different choices of the exponent $\nu$. We employ the masses of different states and their radial wave functions obtained from the study to compute the two-photon and two-gluon decay widths and the E1 and M1 radiative transitions. It is found that the quarkonia mass spectra and the E1 transition can be described by the same interquark model potential of the $CPP_\nu$ with $\nu=1.0$ for $c\bar c$ and  $\nu=0.7$ for $b\bar b$ systems, while the M1 transition (at which the spin of the system changes) and the decay rates in the annihilation channel of quarkonia are better estimated by a shallow potential  with $\nu<1.0$.
\end{abstract}


\end{frontmatter}

% \linenumbers

\section{Introduction}\label{section:intro}
The renaissance in the hadron spectroscopy particularly in the heavy flavour sector associated with the large number of charm and beauty states observed in recent experiments \cite{Aubert1974,Augustin1974,Choi2003,Ecklund2008} has created renewed interest in the study of hadron spectroscopy. Particularly the discovery of $\eta_b(1S)$ state, $\eta_c(2S)$ state and other orbitally excited states in the charmonia and bottonia systems \cite{PDG2008} has stimulated the theoretical phenomenologists to take a new look and refine their parameters that describe the nature of quark-antiquark interaction at the hadronic scale. The goodness of the spectroscopic parameters like the interquark potential and its parameters that describe the masses of the bound states and the corresponding wave functions obtained from the phenomenology, in the descriptions of other properties like the decay (in the annihilation channel) and transition properties now become the prospects for a detail investigation. With the recent CLEO measurements \cite{Ecklund2008,PDG2008} of the two-photon decay rates of the even-parity, P-wave $0^{++}$ ($\chi_{Q0}$) and $2^{++}$ ($\chi_{Q2}$) states and with renewed interest in radiative decays of heavy quarkonium states, it seems appropriate to have another look at the two-photon decay of heavy quarkonia from the view point of quark-antiquark interaction \cite{Lansberg2006,Lansberg2007}. Though the physics of quarkonium decay seems to be better understood within the conventional framework of QCD \cite{Brambilla2005}, unlike the two-photon width of S-wave ($\eta_Q$) which can be predicted from the corresponding $J/\psi$ and $\Upsilon$  leptonic widths, the similar prediction for the P-wave ($\chi_Q$) states is not conclusive. Similarly the radiative transitions in heavy quarkonia ($c\bar c$, $b\bar b$ and $c\bar b$) states have drawn much theoretical interest \cite{Eichten5845,9907240}, as they can provide direct information on the nature of $Q\bar Q$ interaction.\\
Most of the existing theoretical values for the decay rates are based on potential model calculations that employ different types of interquark potentials  \cite{Barbieri1976,Bodwin1992,Gupta1996,Huang1996,Schuler1998,Munz1996,Crater2006,Wang2007,Laverty2009,Godfrey1985,Ebert2003,Barnes1992,Vairo}. We consider the conventional nonrelativistic formalism for computations of the decay properties of the heavy flavour systems. In the traditional non-relativistic bound state calculation, the two-photon and two-gluon widths of quarkonium states depend on the $\ell^{th}$ derivatives of the radial wave function at the origin \cite{Barbieri1976}. As considered by many authors \cite{AKRai2005,AKRai2008,Ebert,Ebert200318,Kim2005}, the contribution from the radiative corrections to these decays are also incorporated in the present study.\\
The paper is organized as follows. In Section \ref{section:cppn}, we describe the phenomenological quark-antiquark interaction potential and extract the parameters that describe the ground state masses of $c\bar c$, $b\bar b$ and $c\bar b$ systems. We also compute the low lying orbital excited states of these systems. In section \ref{section:annihilation} we employ the spectroscopic parameters of the $c\bar c$ and $b\bar b$ systems to study the two photon and two-gluon decay widths. The radiative transitions (E1 and M1) of the $c\bar c$, $b\bar b$ and $c\bar b$ systems are described in section \ref{section:radiative}. In section \ref{section:results} we present, analyze and discuss our results to draw important conclusions.
\section{The phenomenology and extraction of the spectroscopic parameters}\label{section:cppn}
\begin{figure}[!htbp]
\begin{center}
\subfigure{\label{R0CC}
\includegraphics[width=6cm, height=4.5cm, trim=.8cm 1cm 1cm .51cm]{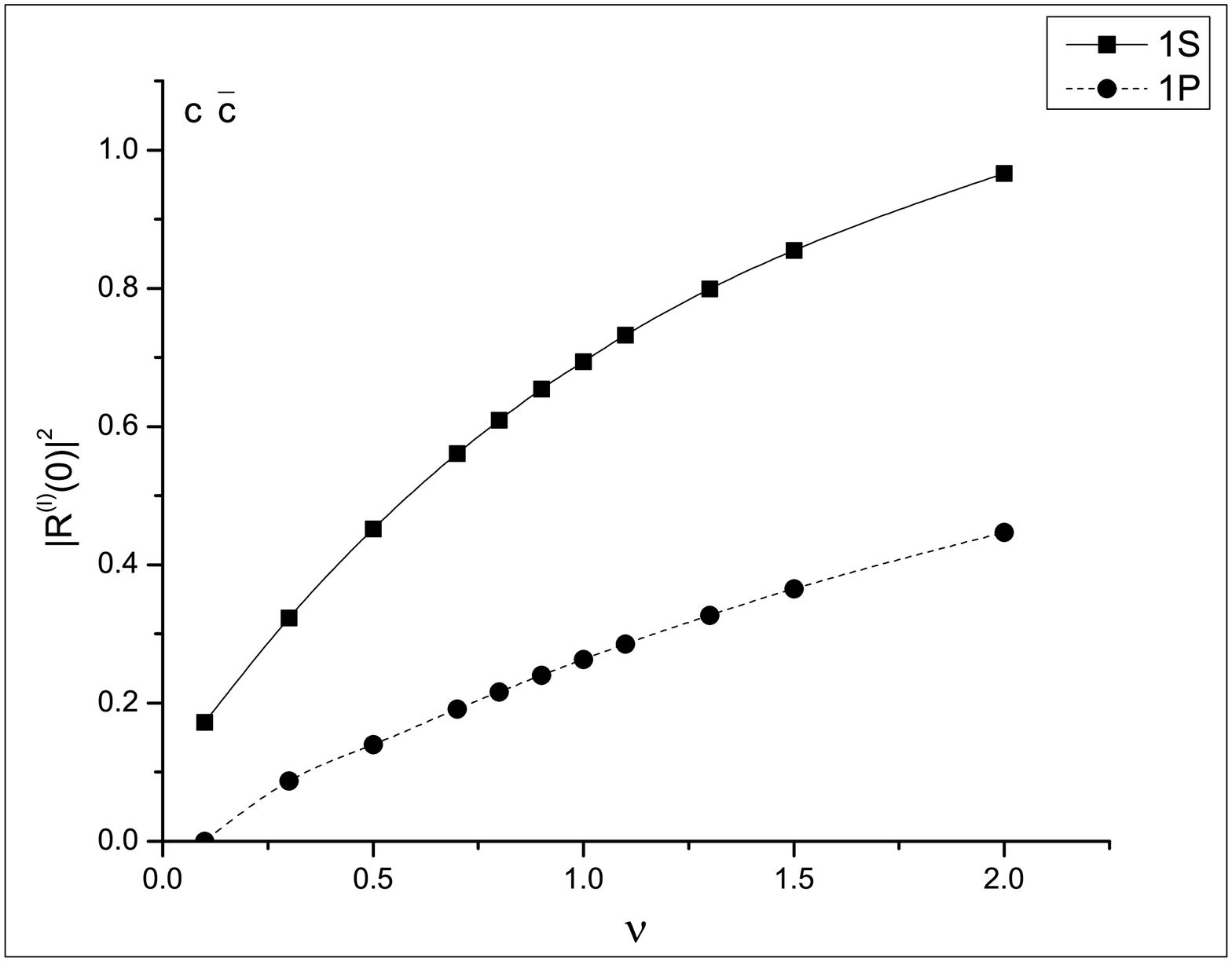}}
\subfigure{\label{R0BB}
\includegraphics[width=6cm, height=4.5cm, trim=.8cm 1cm 1cm .51cm]{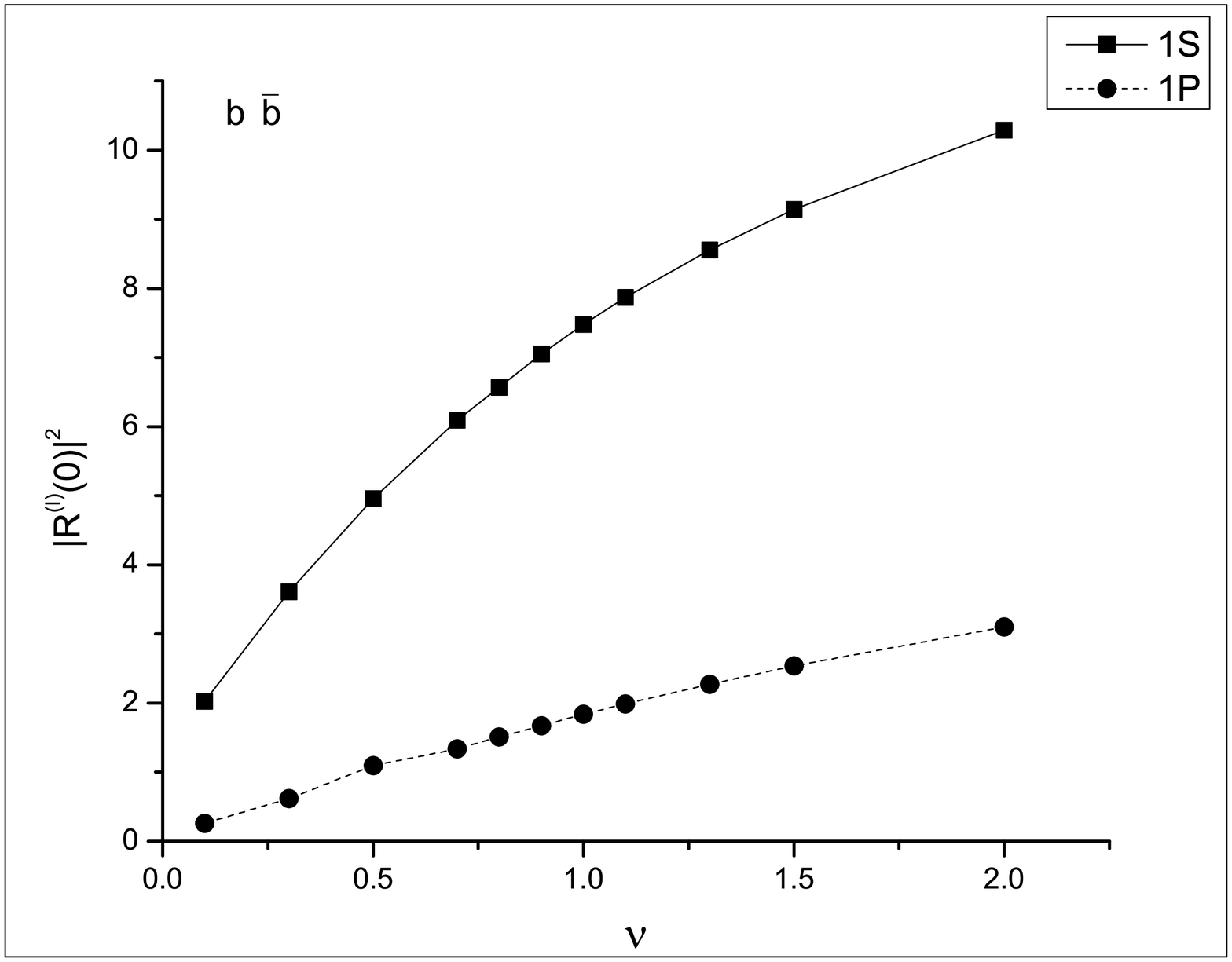}}
\caption{Square of $\ell^{th}$ derivative of radial wave function at origin in $GeV^{2\ell+3}$}\label{R0}
\end{center}
\end{figure}

For the description of the quarkonium bound states, we adopt the phenomenological Coulomb plus
power potential ($CPP_\nu$) expressed as \cite{AKRai2005,AKRai2008}
\begin{equation}\label{4}
V(r)=-\frac{4}{3}\frac{\alpha_s}{r}+Ar^\nu
\end{equation}
Here, $A$ is the confinement strength of the potential and $\alpha_s$ is the
running strong coupling constant which is computed as,
\begin{equation}  \label{alpha}
\alpha _s(\mu^2)=\frac{4 \pi}{(11-\frac{2}{3}n_{f})\ {\rm ln}(\mu^2/ \Lambda^2)}
\end{equation}
Where, $n_f$ is the number of flavors,  $\mu$ is the renormalization
scale related to the constituent quark mass and $\Lambda$ is the QCD scale which is
taken as 0.150 GeV by fixing $\alpha_{s}=0.118$ at the $Z-$boson mass (91 Ge$V$) \cite{PDG2008}.\\
\begin{figure}[!htbp]
\begin{center}
\includegraphics[width=8cm, height=6cm, trim=.8cm 1cm 1cm .51cm]{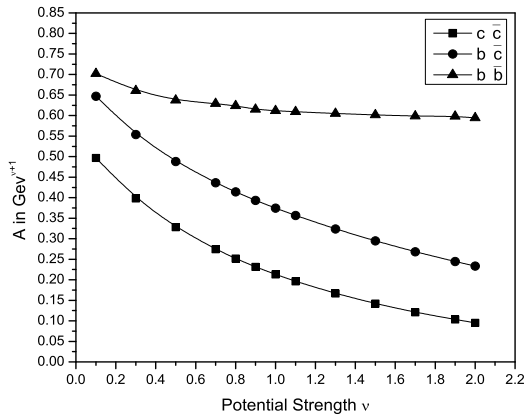}
\caption{Potential strength $A$ (in $GeV^{\nu+1}$) obtained from the ground state spin-average mass against the choices of potential index, $\nu$ ($0.1\leq\nu\leq2.0$)}\label{anu}
\end{center}
\end{figure}
The potential parameter, $A$ of Eqn.\ref{4} is similar to the string strength $\sigma$ of the Cornell potential. We particularly chose to vary $\nu$ in our study as very different interquark potentials can provide fairly good description of the mass spectra, while the transitions and other decay properties are very sensitive to the choice of interquark potential. Thus the present study on the decay properties of heavy flavour mesons based on the $CPP_\nu$ model by varying the exponent $\nu$ ($0.1\leq\nu\leq2.0$) can provide significant understanding of the quark-antiquark interaction in the mesonic states while they undergo a transition or decay through annihilation channels. The different choices of $\nu$ here then correspond to different potential forms. So, the potential parameter $A$ expressed in Ge$V^{\nu+1}$ can be different for each choices of $\nu$. The model potential parameter $A$ and the mass parameter of the quark/antiquark ($m_{1},m_2$) are fixed using the known ground state center of weight (spin average) mass and the hyperfine splitting ($M_{^{3}S_{1}}-M_{^{1}S_{0}}$) of the ground state $c \bar{c}$ and $b \bar{b}$ systems respectively. The spin average mass for the ground state is computed for the different choices of $\nu$ in the range, $0.1 \leq \nu \leq 2.0$. The spin average or the center of weight mass, $M_{CW}$ is calculated from the known experimental/theoretical values of the pseudoscalar ($J=0$) and vector ($J=1$) mesonic mass as
\begin{equation}
M_{n, CW}=\frac{\sum\limits_{J} (2J+1)\ M_{nJ}}{\sum\limits_{J}
(2J+1)}
\end{equation}
The Schr$\ddot{o}$dinger equation is numerically solved using the mathematica notebook [ver. 3.0] of the Runge-Kutta method \cite{rungekutta}.
For computing the mass difference between different spin degenerate
mesonic states, we consider the spin dependent part of the usual one
gluon exchange potential (OGEP) given by
 \cite{Branes2005,Lakhina2006,Voloshin2008,Eichten2008,Gerstein1995}.
Accordingly, the spin-dependent part, $V_{SD}(r)$ contains three types of interaction terms, such as the spin-spin, the spin-orbit and the tensor part as
\begin{eqnarray}\label{spin}
                V_{SD}(r) &=&  V_{SS}(r)\left[S(S+1)-\frac{3}{2}\right]+
                          V_{LS}(r)\left(\vec{L}\cdot\vec{S}\right)+\cr
               && V_{T}(r) \left[S
(S+1)-\frac{3(\vec{S}\cdot\vec{r})(\vec{S}\cdot\vec{r})}{r^2}\right]
              \end{eqnarray}
The spin-orbit term containing $V_{LS}(r)$ and the tensor term containing $V_{T}(r)$ describe the fine structure of the meson states, while the spin-spin term containing $V_{SS}(r)$ proportional to
$2(\vec{s_{q}}\cdot\vec{s_{\bar q}})=S(S+1)-\frac{3}{2} $ gives the spin singlet-triplet hyperfine splitting. The coefficient of these spin-dependent terms of Eqn.\ref{spin} can be written in terms of the vector ($V_V$) and scalar ($V_S$) parts of the static potential, $V(r)$ described in Eqn. \ref{4} as \cite{Voloshin2008}
\begin{equation} \label{LS}
V_{LS}(r)=\frac{1}{2\ m_{1} m_{2}\
r}\left(3\frac{dV_{V}}{dr}-\frac{dV_{S}}{dr}\right)
\end{equation}
\begin{equation} \label{T}
V_{T}(r)=\frac{1}{6\ m_{1} m_{2}
}\left(3\frac{d^{2}V_{V}}{dr^{2}}-\frac{1}{r}\frac{dV_{V}}{dr}\right)
\end{equation}
\begin{equation} \label{SS}
V_{SS}(r)=\frac{1}{3\ m_{1} m_{2}}\nabla^{2} V_{V}=\frac{16 \ \pi
\alpha_{s}}{9\ m_{1} m_{2} } \delta^{(3)}(\vec{r})
\end{equation}
Here $V_V$ is the coulumb part ($1^{st}$ term) and $V_S$ is the confining part ($2^{nd}$ term) of Eqn. \ref{4}.
The computed masses of the $Q\bar Q(^{2s+1}L_J)$ states are listed in Table \ref{masscc}, \ref{massbc}
and \ref{massbb} for different combinations of $Q\in b,c$. The spectroscopic parameters thus correspond to the fitted quark masses, the potential strength $A$, the potential exponent, $\nu$ and the corresponding radial wave functions. The fitted mass parameters are $m_c=1.28$ $GeV/c^2$, $m_b=4.4$ $GeV/c^2$ while the potential strength $A$ for each choices of $\nu$ are shown in Fig. \ref{anu}. The $\ell^{th}$ derivative of the corresponding radial wave functions at the origin obtained for each choices of the exponent $\nu$ are plotted in Fig. \ref{R0}.
\begin{table}[htbp]
\begin{center}
\caption{Masses (in $GeV$) of charmonium states ($n^{2S+1}L_J$) in the $CPP_\nu$ mode with different choices of $\nu$}\label{masscc}
\begin{tabular}{rrrrrrrrrr}
\hline
&\multicolumn{8}{c}{$\nu$}&Expt.\\
\cline{2-9}
State&0.1&0.5&0.7&0.9&1.0&1.1&1.5&2.0&\cite{PDG2008}\\
\hline
$1^3S_1$&3.076&3.088&3.093&3.097&3.099&3.100&3.106&3.111&3.097\\
$1^1S_0$&3.045&3.008&2.994&2.981&2.976&2.971&2.955&2.940&2.980\\
$1^3P_2$&3.174&3.357&3.430&3.495&3.524&3.552&3.647&3.737&3.556\\
$1^3P_1$&3.171&3.347&3.419&3.484&3.514&3.542&3.640&3.736&3.511\\
$1^3P_0$&3.167&3.324&3.386&3.441&3.466&3.489&3.570&3.647&3.415\\
$1^1P_1$&3.172&3.350&3.422&3.485&3.514&3.542&3.636&3.727&3.526\\
$1^3D_3$&3.215&3.520&3.653&3.774&3.830&3.830&4.068&4.249&\\
$1^3D_2$&3.214&3.523&3.662&3.792&3.854&3.914&4.129&4.354&\\
$1^3D_1$&3.214&3.521&3.663&3.796&3.860&3.923&4.152&4.399&3.770\\
$1^1D_2$&3.214&3.521&3.658&3.785&3.844&3.901&4.105&4.314&\\
$2^3P_2$&3.121&3.461&3.633&3.803&3.870&3.970&4.292&4.661&3.929\\
$2^3P_1$&3.120&3.455&3.624&3.792&3.875&3.958&4.277&4.647&\\
$2^3P_0$&3.118&3.439&3.599&3.757&3.835&3.912&4.206&4.541&\\
$2^1P_1$&3.120&3.457&3.626&3.740&3.877&3.960&4.277&4.643&\\
$2^3S_1$&3.087&3.326&3.444&3.558&3.615&3.670&3.881&4.119&3.686\\
$2^1S_0$&3.079&3.290&3.390&3.487&3.533&3.579&3.749&3.936&3.638\\
\hline
\end{tabular}
\end{center}
\end{table}

\begin{table}[htbp]
\begin{center}
\caption{Masses (in $GeV$) of $B_c$ states ($n^{2S+1}L_J$) in the $CPP_\nu$ mode with different choices of $\nu$}\label{massbc}
\begin{tabular}{rrrrrrrrrrrr}
\hline
&\multicolumn{6}{c}{$\nu$}&\multicolumn{5}{c}{Others}\\
\cline{2-12}
State&0.1&0.5&0.7&1.0&1.5& 2.0&\cite{Vary}&\cite{Eq7}&\cite{GLKT}&\cite{Ebert}&\cite{Godfrey}\\
\hline
$1^3S_1$&6.326&6.335&6.339&6.343&6.348&6.351&6.416&6.337&6.317&6.332&6.338\\
$1^1S_0$&6.305&6.279&6.268&6.256&6.241&6.231&6.380&6.264&6.253&6.270&6.271\\
$1^3P_2$&6.453&6.679&6.771&6.889&7.043&7.156&6.837&6.747&6.743&6.762&6.768\\
$1^3P_1$&6.451&6.673&6.764&6.882&7.038&7.155&6.772&6.730&6.717&6.734&6.741\\
$1^3P_0$&6.448&6.656&6.740&6.848&6.989&7.092&6.693&6.700&6.683&6.699&6.706\\
$1^1P_1$&6.452&6.675&6.765&6.882&7.035&7.148&6.775&6.736&6.729&6.749&6.750\\
$1^3D_3$&6.505&6.889&7.059&7.286&7.599&7.840&7.003&7.005&7.007&7.081&7.045\\
$1^3D_2$&6.504&6.891&7.065&7.304&7.643&7.915&7.000&7.012&7.001&7.077&7.041\\
$1^3D_1$&6.504&6.890&7.066&7.308&7.659&7.947&6.959&7.012&7.008&7.720&7.028\\
$1^1D_2$&6.505&6.890&7.062&7.297&7.626&7.886&7.001&7.009&7.016&7.079&7.036\\
$2^3P_2$&6.387&6.812&7.027&7.345&7.852&8.313&7.186&7.153&7.134&7.156&7.164\\
$2^3P_1$&6.386&6.808&7.021&7.337&7.843&8.303&7.136&7.135&7.113&7.126&7.145\\
$2^3P_0$&6.384&6.797&7.003&7.309&7.792&8.229&7.081&7.108&7.088&7.091&7.122\\
$2^1P_1$&6.386&6.809&7.022&7.339&7.842&8.301&7.139&7.142&7.124&7.145&7.150\\
$2^3S_1$&6.343&6.640&6.785&6.997&7.326&7.617&6.896&6.899&6.902&6.881&6.887\\
$2^1S_0$&6.338&6.615&6.748&6.939&7.232&7.487&6.875&6.856&6.867&6.835&6.855\\
\hline
\end{tabular}
\end{center}
\end{table}

\begin{table}[htbp]
\begin{center}
\caption{Masses (in $GeV$) of bottonia states ($n^{2S+1}L_J$) in the $CPP_\nu$ mode with different choices of $\nu$}\label{massbb}
\begin{tabular}{rrrrrrrrr}
\hline
&\multicolumn{7}{c}{$\nu$}&Expt.\\
\cline{2-8}
State&0.1&0.5&0.7&0.9&1.0&1.5&2.0&\cite{PDG2008}\\
\hline
$1^3S_1$&9.447&9.455&9.458&9.460&9.461&9.466&9.468&9.460\\
$1^1S_0$&9.427&9.404&9.395&9.388&9.385&9.372&9.363&9.389\\
$1^3P_2$&9.595&9.838&9.938&10.025&10.065&10.231&10.356&9.912\\
$1^3P_1$&9.593&9.831&9.929&10.016&10.056&10.223&10.348&9.893\\
$1^3P_0$&9.590&9.817&9.909&9.989&10.026&10.179&10.292&9.859\\
$1^1P_1$&9.594&9.834&9.932&10.018&10.057&10.223&10.346&\\
$1^3D_3$&9.653&10.066&10.250&10.419&10.498&10.842&11.111&\\
$1^3D_2$&9.653&10.067&10.254&10.427&10.509&10.871&11.163&10.162\\
$1^3D_1$&9.652&10.066&10.253&10.428&10.511&10.881&11.182&\\
$1^1D_2$&9.653&10.060&10.252&10.423&10.504&10.859&11.143&\\
$2^3P_2$&9.530&9.984&10.214&10.443&10.556&11.100&11.599&10.269\\
$2^3P_1$&9.529&9.970&10.208&10.434&10.547&11.088&11.583&10.255\\
$2^3P_0$&9.528&9.970&10.130&10.413&10.522&11.044&11.518&10.232\\
$2^1P_1$&9.530&9.981&10.210&10.437&10.549&11.090&11.585&\\
$2^3S_1$&9.481&9.796&9.950&10.101&10.175&10.524&10.835&10.023\\
$2^1S_0$&9.476&9.773&9.918&10.057&10.125&10.444&10.724&\\
\hline
\end{tabular}
\end{center}
\end{table}
\section{Two-photon and two-gluon decay widths for quarkonium states}\label{section:annihilation}
The extracted model parameters and the radial wavefunctions are being employed here to compute the two-photon ($\Gamma_{\gamma\gamma}(Q\bar Q)$) and two gluon ($\Gamma_{gg}(Q\bar Q)$) decay widths. Two photon widths of orbitally excited quarkonium states  $\chi_{QJ=0,2}\rightarrow\gamma\gamma$ are suppressed by the mass of the heavy quark while the two-photon decays of the spin one state $\chi_{Q,1}$ is forbidden by the Landau-Yang theorem \cite{Landau1949,Yang1950}. Most of the quark model predictions \cite{PDG2008,Ebert2003,Zou2003} for the $\eta_Q\rightarrow\gamma\gamma$ width are comparable with the experimental value. However the theoretical predictions for the $P$-wave ($\chi_{Q0,2}\rightarrow\gamma\gamma$) widths differ largely from the experimental observations \cite{PDG2008}. This warranted to incorporate contribution from QCD corrections. Thus with the one-loop QCD radiative corrections the decay widths of $^1S_0$ ($\eta_Q$), $^3P_0$ ($\chi_{Q0}$) and $^3P_2$ ($\chi_{Q2}$) states into two photons are computed according to the non relativistic expression given by  \cite{Barbieri1976,Huang1996,AKRai2005,AKRai2008,Bhavin2009,Petrelli1998,Kwong1988}
\begin{equation}\label{1}
\Gamma_{\gamma\gamma}(\eta_Q)=\frac{3Q_e^4\alpha_{em}^2M_{\eta_Q}|R_0(0)|^2}{2m_Q^3}\left[1-\frac{\alpha_s}{\pi}\frac{(20-\pi^2)}{3}\right]
\end{equation}
\begin{equation}\label{2}
\Gamma_{\gamma\gamma}(\chi_{Q0})=\frac{27Q_e^4\alpha_{em}^2M_{\chi_{Q0}}|R_1^{'}(0)|^2}{2m_Q^5}\left[1+B_0\frac{\alpha_s}{\pi}\right]
\end{equation}
\begin{equation}\label{3}
\Gamma_{\gamma\gamma}(\chi_{Q2})=\frac{4}{15}\frac{27Q_e^4\alpha_{em}^2M_{\chi_{Q2}}|R_1^{'}(0)|^2}{2m_Q^5}\left[1+B_2\frac{\alpha_s}{\pi}\right]
\end{equation}
Here, $B_0=\pi^2/3-28/9$ and $B_2=-16/3$ are the next to leading order (NLO) QCD radiative corrections \cite{Kwong1988,Barbieri1981,Mangano1995} and are considered to be the same for charmonium and bottonium systems.\\
Similarly, the two-gluon decay width of $\eta_Q$, $\chi_{Q0}$ and $\chi_{Q2}$  states are given by \cite{Lansberg2009},
\begin{equation}\label{gamma1}
\Gamma_{gg}(\eta_{Q})=\frac{\alpha_s^2M_{\eta_c}|R_0(0)|^2}{3m_Q^3}[1+C_Q(\alpha_s/\pi)]
\end{equation}
\begin{equation}\label{gamma2}
\Gamma_{gg}(\chi_{Q0})=\frac{3\alpha_s^2M_{\chi_{Q0}}|R_1^{'}(0)|^2}{m_Q^5}[1+C_{0Q}(\alpha_s/\pi)]
\end{equation}
\begin{equation}\label{gamma3}
\Gamma_{gg}(\chi_{Q2})=\left(\frac{4}{15}\right)\frac{3\alpha_s^2M_{\chi_{Q2}}|R_1^{'}(0)|^2}{m_Q^5}[1+C_{2Q}(\alpha_s/\pi)]
\end{equation}
Here, the quantities in the square brackets are the NLO QCD radiative corrections  \cite{Kwong1988,Barbieri1981,Mangano1995} and the coefficients, $C_Q$, $C_{0Q}$ and $C_{2Q}$ for the charmonia $(Q=c)$ and bottonia ($Q=b$) are listed in Table \ref{nlo}.\\
\begin{table}[h]tbp
\begin{center}
\caption{NLO QCD Correction coefficients for the two-gluon decays}\label{nlo}
\begin{tabular}{rrr}
\hline
&Q=c&Q=b\\
\hline
$C_Q$&4.8&4.4\\
$C_{0Q}$&8.77&10.0\\
$C_{2Q}$&$-4.827$&$-0.1$\\
\hline
\end{tabular}
\end{center}
\end{table}
The value of the radial wave functions $|R_{Q\bar Q}^{(\ell)}(0)|^2$ of Eqns. \ref{1}-\ref{3} are obtained from the numerical solution of the schrodinger equation corresponding to the $CPP_\nu$ potential model description of the $Q\bar Q$ bound states described in section \ref{section:cppn}. The computed digamma widths of the $c\bar c$ and $b\bar b$ states are listed in Table \ref{twophotoncc} and \ref{twophotonbb} while the digluon widths of $c\bar c$ and $b\bar b$ states are listed against the exponent, $\nu$ in Table \ref{twogluoncc} and \ref{twogluonbb} respectively.
\begin{table}[htbp]\label{t3}
\begin{center}
\caption{Two-photon decay widths (in $keV$) of $c\bar c$ states with different choices of $\nu$}\label{twophotoncc}
\begin{tabular}{crrr}
\hline
$\nu$&$\Gamma_{\gamma\gamma}(\eta_{c})$&$\Gamma_{\gamma\gamma}(\chi_{c0})$&$\Gamma_{\gamma\gamma}(\chi_{c2})$\\
\hline
0.5&    6.375&  2.539&  0.309\\
0.7&    7.875&  4.839&  0.529\\
1.0&    9.684&  9.357&  1.162\\
\hline
\cite{Ecklund2008}&-&$2.36\pm0.35\pm0.22$&$0.66\pm0.07\pm0.06$\\
\cite{PDG2008}  &$6.41^{+3.657}_{-3.123}$&$2.397\pm0.4$ &$0.493\pm0.06$\\
\cite{Huang1996}    &-&3.72 &0.49\\
\cite{Schuler1998}  &-&2.50 &0.28\\
\cite{Munz1996} &-&1.39 &0.44\\
\cite{Crater2006}   &-&3.34 &0.43\\
\cite{Wang2007} &-&3.78 &-  \\
\cite{Laverty2009}  &-&1.99 &0.30\\
\cite{Godfrey1985}  &-&1.29 &0.46\\
\cite{Ebert2003}    &-&2.90 &0.50\\
\cite{Barnes1992}   &-&1.56 &0.56\\
\cite{Ebert}  &4.33&-&    -\\
\cite{Barbieri1981}&- &3.50 &0.93\\
\cite{Floriana}&4.252&-&-\\
\hline
\end{tabular}
\end{center}
\end{table}

\begin{table}[htbp]
\begin{center}
\caption{Two-photon decay widths (in $keV$) of $b\bar b$ states with different choices of $\nu$}\label{twophotonbb}
\begin{tabular}{crrr}
\hline
$\nu$&$\Gamma_{\gamma\gamma}(\eta_{b})$&$\Gamma_{\gamma\gamma}(\chi_{b0})$&$\Gamma_{\gamma\gamma}(\chi_{b2})$\\
\hline
0.5&    0.410&  0.064&  0.011\\
0.7&    0.504&  0.097&  0.016\\
1.0&    0.618&  0.185&  0.030\\
\hline
\cite{Gupta1996}&-&0.080&0.008\\
\cite{Schuler1998}&0.460&-&-\\
\cite{Munz1996}&$0.220\pm0.040$&-&-\\
\cite{Laverty2009}pert.&0.30&0.032&0.007\\
\cite{Laverty2009}nonpert.&0.32&0.094&0.005\\
\cite{Godfrey1985}&0.214&-&-\\
\cite{Ebert200318}&0.350&-&-\\
\cite{Kim2005}&$0.384\pm0.047$&-&-\\
\cite{Lakhina2006}&0.230&-&-\\
\cite{Floriana}&0.313&-&-\\
\cite{Fabiano2003}&$0.466\pm0.101$&-&-\\
\cite{Ackleh1992}&0.170&-&-\\
\cite{Ahmady1995}&0.520&-&-\\
\cite{lansbergpham}&0.560&-&-\\
\hline
\end{tabular}
\end{center}
\end{table}

\begin{table}[t]
\caption{Two-gluon decay widths (in $MeV$) of $c\bar c$ states with different choices of $\nu$}\label{twogluoncc}
\begin{center}
\begin{tabular}{crrr}
\hline
$\nu$&$\Gamma_{gg}(\eta_{c})$&$\Gamma_{gg}(\chi_{c0})$&$\Gamma_{gg}(\chi_{c2})$\\
\hline
0.5&    32.209& 10.467& 1.169\\
0.7&    39.790& 19.949& 2.003\\
1.0&    48.927& 38.574& 4.396\\
\hline
\cite{PDG2008}&$26.7\pm3.0$&$10.2\pm0.7$&$2.03\pm0.12$\\
\cite{Laverty2009}pert.&15.70&4.68&1.72\\
\cite{Laverty2009}nonpert.&10.57&4.88&0.69\\
\hline
\end{tabular}
\end{center}
\end{table}

\begin{table}[htbp]
\begin{center}
\caption{Two-gluon decay widths (in $MeV$) of $b\bar b$ states with different choices of $\nu$}\label{twogluonbb}
\begin{tabular}{crrr}
\hline
$\nu$&$\Gamma_{gg}(\eta_{b})$&$\Gamma_{gg}(\chi_{b0})$&$\Gamma_{gg}(\chi_{b2})$\\
\hline
0.5&11.921& 1.826&  0.285\\
0.7&14.644& 2.745&  0.429\\
1.0&17.945& 5.250&  0.822\\
\hline
\cite{Gupta1996}&12.46&2.15&0.22\\
\cite{Laverty2009}pert.&11.49&0.96&0.33\\
\cite{Laverty2009}nonpert.&12.39&2.74&0.25\\
\hline
\end{tabular}
\end{center}
\end{table}
\section{Radiative transitions}\label{section:radiative}
As the mass spectra of the quarkonia states are well known, the investigation of radiative transitions become important and such study could help us to understand the theory of strong interaction in the nonperturbative regime of QCD. The motivation for the present work is to study the radiative transitions (E1 and M1 transitions) of the $Q\bar Q$ $(Q\in b,c)$ states and to investigate the interquark potentials that provides the right description of the quarkonia states. The nonrelativistic treatment adopted for the study of these heavy flavour mesonic systems allows as to apply the usual multipole expansion in electrodynamics to compute the transition between the quarkonia states with the emission of a photon. The leading terms in this expansion correspond to the E1 and M1 transitions. The electric dipole term (E1) is responsible for the transition between the $S$ and $P$ states without changing the spin of the quark-antiquark pair, while the magnetic dipole term (M1) describes the transition between $S=1$ and $S=0$ states without changing relative orbital momentum ($L$) of the quark-antiquark pair. Accordingly, electric transitions do not change quark spin. These transitions have $\Delta L=\pm1$ and $\Delta S=0$. The E1 radiative transition width between initial state, ($n^{2s+1}L_J$) to final state, ($n^{'2s+1}L{'}_J^{'}$) is given by \cite{Eichten3090},
\begin{equation}
\Gamma(i\rightarrow f+\gamma)=\frac{4\alpha\langle e_Q\rangle^2}{3}(2J^{'}+1)S_{if}^E\omega^3|\varepsilon_{if}|^2
\end{equation}
Here, $\alpha=1/137$,  $\omega$ is the photon energy and $\langle e_Q\rangle$, the mean charge content of the $Q\bar Q$ system are expressed as
\begin{equation}
\omega=\frac{M_i^2-M_f^2}{2M_i}
\end{equation}
and
\begin{equation}
\langle e_Q\rangle=\left|\frac{m_{\bar Q}e_Q-m_Qe_{\bar Q}}{m_Q+m_{\bar Q}}\right|
\end{equation}
respectively. Here, $M_i$ and $M_f$ are the initial and final state mass of the quarkonia respectively.
The statistical factor $S_{if}^E=S_{fi}^E$ is given by,
\begin{equation}
S_{if}^E=max(\ell,\ell^{'})\left\{ \begin{array}{ccc}
J&1&J^{'}\\
\ell^{'}&s&\ell\\\end{array} \right\}^2
\end{equation}
The overlap integral $\varepsilon_{if}$ is given by
\begin{equation}
\varepsilon_{if}=\frac{3}{\omega}\int_0^\infty dru_{n\ell}(r)u_{n^{'}\ell^{'}(r)}\left[\frac{\omega r}{2}j_0\left(\frac{\omega r}{2}\right)-j_1\left(\frac{\omega r}{2}\right)\right]
\end{equation}
The E1 transition rates of charmonia, bottonia and $B_c$ systems are listed in Table \ref{E1cc}, \ref{E1bb} and \ref{E1bc}.
\begin{table}
\begin{center}
\caption{E1 transition widths (in $keV$) of $c\bar c$ states against potential index $\nu$}\label{E1cc}
\begin{tabular}{ccrrrrr}
\hline
Initial&Final&\multicolumn{3}{c}{$\nu$}&\multicolumn{2}{c}{Others}\\
\cline{3-7}
State&State&0.5&0.7&1.0&\cite{PDG2008}&\cite{0412158}\\
\hline
$1^3P_2$&$1^3S_1$&163.98&250.22&383.48&406.00&315\\
$1^3P_1$&$1^3S_1$&148.31&229.62&361.21&320.40&41\\
$1^3P_0$&$1^3S_1$&115.53&173.42&263.62&130.56&120\\
$1^1P_1$&$1^1S_0$&302.44&451.50&671.05&&482\\
$1^3D_3$&$1^3P_2$&137.86&245.87&432.36&&402\\
$1^3D_2$&$1^3P_2$&36.23&68.29&131.01&&69.5\\
$1^3D_2$&$1^3P_1$&127.47&231.43&423.43&&31.3\\
$1^3D_1$&$1^3P_2$&3.89&7.68&15.23&&3.88\\
$1^3D_1$&$1^3P_1$&68.65&129.97&245.70&&99\\
$1^3D_1$&$1^3P_0$&127.71&240.03&447.72&&299\\
$1^1D_2$&$1^1P_1$&157.16&285.76&523.89&&389\\
$2^3P_2$&$1^3S_1$&70.34&140.10&252.68&&\\
$2^3P_2$&$2^3S_1$&36.97&83.93&164.41&&\\
$2^3P_1$&$1^3S_1$&66.96&132.92&258.15&&\\
$2^3P_1$&$2^3S_1$&32.65&73.81&172.63&&\\
$2^3P_0$&$1^3S_1$&58.48&114.28&216.53&&\\
$2^3P_0$&$2^3S_1$&22.63&49.36&112.39&&\\
$2^3P_1$&$1^1S_0$&122.85&226.91&415.25&&\\
$2^1P_1$&$2^1S_0$&65.04&146.99&332.99&&\\
\hline
\end{tabular}
\end{center}
\end{table}

\begin{table}
\begin{center}
\caption{E1 transition widths (in $keV$) of $b\bar b$ states against potential index $\nu$}\label{E1bb}
\begin{tabular}{ccrrrrr}
\hline
Initial&Final&\multicolumn{3}{c}{$\nu$}&\multicolumn{2}{c}{Others}\\
\cline{3-7}
State&State&0.5&0.7&1.0&\cite{PDG2008}&\cite{0412158}\\
\hline
$1^3P_2$&$1^3S_1$&29.06&44.28&70.29&11.88&31.6\\
$1^3P_1$&$1^3S_1$&27.58&42.00&67.43&18.91&27.8\\
$1^3P_0$&$1^3S_1$&24.75&37.18&58.42&3.24&22.2\\
$1^1P_1$&$1^1S_0$&40.29&60.45&94.05&&\\
$1^3D_3$&$1^3P_2$&17.28&41.00&76.82&&22.6\\
$1^3D_2$&$1^3P_2$&4.37&10.63&20.60&&\\
$1^3D_2$&$1^3P_1$&14.32&34.53&65.35&&\\
$1^3D_1$&$1^3P_2$&0.48&1.17&2.32&&0.50\\
$1^3D_1$&$1^3P_1$&7.86&19.02&36.75&&10.7\\
$1^3D_1$&$1^3P_0$&12.38&30.04&58.47&&20.1\\
$1^1D_2$&$1^1P_1$&16.84&44.06&83.96&&\\
$2^3P_2$&$1^3S_1$&11.05&20.42&35.84&&12.7\\
$2^3P_2$&$2^3S_1$&5.54&13.34&32.92&&14.5\\
$2^3P_1$&$1^3S_1$&10.19&19.92&34.86&&12.0\\
$2^3P_1$&$2^3S_1$&4.43&12.50&30.82&&12.4\\
$2^3P_0$&$1^3S_1$&10.19&14.14&32.24&&10.9\\
$2^3P_0$&$2^3S_1$&4.43&4.44&25.40&&9.17\\
$2^1P_1$&$1^1S_0$&14.39&25.84&43.99&&\\
\hline
\end{tabular}
\end{center}
\end{table}

\begin{table}
\begin{center}
\caption{E1 transition widths (in $keV$) of $B_c$ states against potential index $\nu$}\label{E1bc}
\begin{tabular}{ccrrrrrrr}
\hline
Initial&Final&\multicolumn{3}{c}{$\nu$}&\multicolumn{4}{c}{Others}\\
\cline{3-9}
State&State&0.5&0.7&1.0&\cite{Godfrey}&\cite{Ebert}&\cite{Vary}&\cite{Eq7}\\
\hline
$1^3P_2$&$1^3S_1$&82.04&128.18&200.07&83&107&109.8&112.6\\
$1^3P_1$&$1^3S_1$&78.22&122.73&193.48&11&78.9&67.8&99.5\\
$1^3P_0$&$1^3S_1$&67.93&105.03&163.13&60&67.2&32.1&79.2\\
$1^1P_1$&$1^1S_0$&119.47&184.66&282.73&&&81.8&56.4\\
$1^3D_3$&$1^3P_2$&68.56&125.28&225.588&&&18.7&98.7\\
$1^3D_2$&$1^3P_2$&17.6&33.13&63.37&&&4.4&24.7\\
$1^3D_2$&$1^3P_1$&57.06&105.92&198.57&&&35.8&88.8\\
$1^3D_1$&$1^3P_2$&1.93&3.71&7.22&&&0.2&2.7\\
$1^3D_1$&$1^3P_1$&31.3&59.38&113.06&&&11.4&49.3\\
$1^3D_1$&$1^3P_0$&51.4&97.17&183.53&&&43.9&88.6\\
$1^1D_2$&$1^1P_1$&73.20&136.21&325.46&&&120.8&92.5\\
$2^3P_2$&$1^3S_1$&34.78&68.56&134.25&&&9.1&25.8\\
$2^3P_2$&$2^3S_1$&17.88&41.63&97.24&&&91.3&73.8\\
$2^3P_1$&$1^3S_1$&33.89&66.66&130.62&&&6.3&22.1\\
$2^3P_1$&$2^3S_1$&16.74&38.91&91.62&&&53.5&54.3\\
$2^3P_0$&$1^3S_1$&31.53&61.18&118.43&&&5.9&21.9\\
$2^3P_0$&$2^3S_1$&13.86&31.36&73.25&&&24.3&41.2\\
$2^1P_1$&$1^1S_0$&48.08&91.81&174.50&&&&\\
$2^1P_1$&$2^1S_0$&24.87&57.80&137.46&&&&\\
\hline
\end{tabular}
\end{center}
\end{table}
The magnetic dipole transitions (M1) flip the quark spin, so their amplitudes are proportional to the quark magnetic moments and thus related inversely to the quark mass. Thus, M1 transitions are weaker than the E1 transitions and this causes difficulties in the experimental observations. Along with other exclusive processes, the magnetic dipole (M1) transitions from the spin-triplet S-wave vector (V) state to the spin-singlet S-wave pseudoscalar (P) state have also been considered as a valuable testing ground to further constrain the phenomenological quark model of hadrons. The M1 transitions have $\Delta L=0$ and the $n^{2s+1}L_J\rightarrow n^{'2s{'}+1}L_{J^{'}}+\gamma$ transition rate for $Q\bar Q$ mesonic system is given by \cite{Radford,0412158,Lahde}
\begin{equation}\label{eq.M1}
\Gamma(i\rightarrow f+\gamma)=\frac{\alpha}{3}\mu^2\omega^3S_{if}^M(2J^{'}+1)M^2
\end{equation}
Where
\begin{equation}
\mu=\left|\frac{e_Q}{m_Q}-\frac{e_{\bar Q}}{m_{\bar Q}}\right|
\end{equation}
and, $m_{Q/\bar Q}$ is the mass of the heavy quark/antiquark. The statistical factor $S_{if}^M=S_{fi}^M$ is
\begin{equation}
S_{if}^M=6(2s+1)(2s^{'}+1)\left\{ \begin{array}{ccc}
J&1&J^{'}\\s^{'}&\ell&s\end{array} \right\}^2\left\{ \begin{array}{ccc}
1&\frac{1}{2}&\frac{1}{2}\\\frac{1}{2}&s^{'}&s\end{array} \right\}^2
\end{equation}
The matrix element $M$ contains the integral related to the static term ($I_1$) and the term related to the contribution from the two quark confining exchange current ($I_c$) in the non-relativistic limit as
\begin{equation}
M=I_1+I_c
\end{equation}
Where
\begin{equation}
I_1=\langle f|j_0(\omega r/2)|i\rangle
\end{equation}
and
\begin{equation}
I_c =-\frac{1}{m_Q}\langle f|Ar^\nu|i\rangle
\end{equation}
Here, $Ar^\nu$ is the confining part of the quark-antiquark potential described by $CPP_\nu$ model.\\
The magnetic dipole transition rate (allowed M1 transitions) corresponds to triplet-singlet transitions between $S$- wave states, $\Gamma(n^3S_1\rightarrow n^1S_0+\gamma)$ of the $c\bar c$, $b\bar b$ and $c\bar b$ systems are listed in Table \ref{M1cc}, \ref{M1bb} and \ref{M1bc} respectively. Other allowed M1 transitions $n^3S_1\rightarrow n^{'1}S_0+\gamma$ with $n>n^{'}$ are nonrelativistically forbidden \cite{PRD.30.1924} and require relativistic treatment which is beyond the scope of the present study. The values within the brackets are the results computed by incorporating the two-quark exchange current contribution as given by \cite{Lahde,Lahde645}. Other theoretical model predictions and available experimental results are also listed for comparison.
\begin{table}
\begin{center}
\caption{M1 transition widths (in $keV$) of $c\bar c$ states against potential index $\nu$}\label{M1cc}
\begin{tabular}{ccccccccc}
\hline
Initial&Final&\multicolumn{4}{c}{$\nu$}&\multicolumn{3}{c}{Others}\\
\cline{3-9}
State&State&0.5&0.7&0.8&1.0&\cite{0412158}&\cite{Ebert}&\cite{Lahde645}\\
\hline
$1^3S_1$&$1^1S_0$&1.29(0.47)&2.41(0.96)&3.12(1.28)&4.57(2.01)&2.0&1.05&1.05\\
$2^3S_1$&$2^1S_0$&0.12(0.02)&0.40(0.07)&0.61(0.10)&1.39(0.20)&&1.26&0.14\\
$3^3S_1$&$3^1S_0$&0.04(0.004)&0.16(0.009)&0.28(0.011)&0.76(0.012)&&&0.0124\\
\hline
\end{tabular}
\end{center}
Expt \cite{PDG2008}: $\Gamma(1^3S_1\rightarrow1^1S_0\gamma)=1.21\pm0.37$
\end{table}

\begin{table}
\begin{center}
\caption{M1 transition widths (in $eV$) of $b\bar b$ states against potential index $\nu$}\label{M1bb}
\begin{tabular}{ccccccccc}
\hline
Initial&Final&\multicolumn{4}{c}{$\nu$}&\multicolumn{3}{c}{Others}\\
\cline{3-9}
State&State&0.5&0.7&0.8&1.0&\cite{0412158}&\cite{Ebert}&\cite{Lahde645}\\
\hline
$1^3S_1$&$1^1S_0$&7.28(5.26)&13.70(10.11)&16.47(12.22)&24.01(18.20)&8.95&9.7&18.9\\
$2^3S_1$&$2^1S_0$&0.67(0.41)&1.80(1.10)&3.02(1.80)&6.87(4.01)&1.51&1.6&2.77\\
$3^3S_1$&$3^1S_0$&0.19(0.10)&0.76(0.38)&1.34(0.64)&1.98(0.86)&0.83&0.9&1.32\\
\hline
\end{tabular}
\end{center}
\end{table}

\begin{table}
\begin{center}
\caption{M1 transition widths (in $keV$) of $B_c$ states against potential index $\nu$}\label{M1bc}
\begin{tabular}{ccccccrrrc}
\hline
Initial&Final&\multicolumn{4}{c}{$\nu$}&\multicolumn{4}{c}{Others}\\
\cline{3-10}
State&State&0.5&0.7&0.8&1.0&\cite{Ebert}&\cite{Gershtein3613}&\cite{Godfrey}&\cite{Lahde645}\\
\hline
$1^3S_1$&$1^1S_0$&0.15(0.02)&0.30(0.06)&0.37(0.08)&0.55(0.13)&0.07&0.01&0.08&0.05\\
$2^3S_1$&$2^1S_0$&0.01(0.01)&0.04(0.06)&0.07(0.15)&0.17(0.48)&0.03&0.03&0.01&0.004\\
$3^3S_1$&$3^1S_0$&0.005(0.01)&0.02(0.07)&0.03(0.18)&0.09(0.79)&&&&0.08 eV\\
\hline
\end{tabular}
\end{center}
\end{table}
\section{Results and Discussions}\label{section:results}
\begin{figure}[!htbp]
\begin{center}
\subfigure{\label{fig:sdmasscc}
\includegraphics[width=5cm, height=4cm, trim=.8cm 1cm 1cm .51cm]{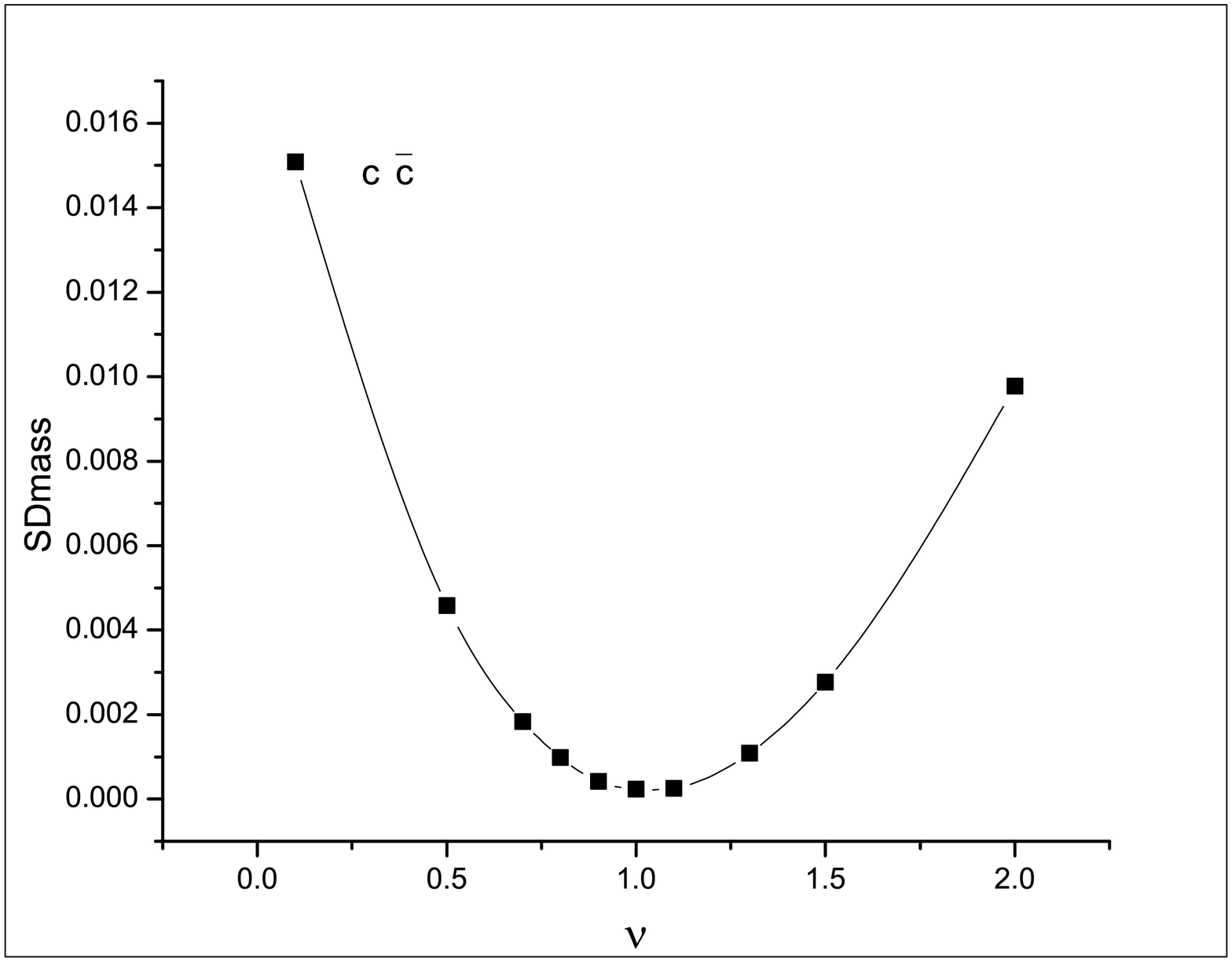}}
\subfigure{\label{fig:sdmassbb}
\includegraphics[width=5cm, height=4cm, trim=.8cm 1cm 1cm .51cm]{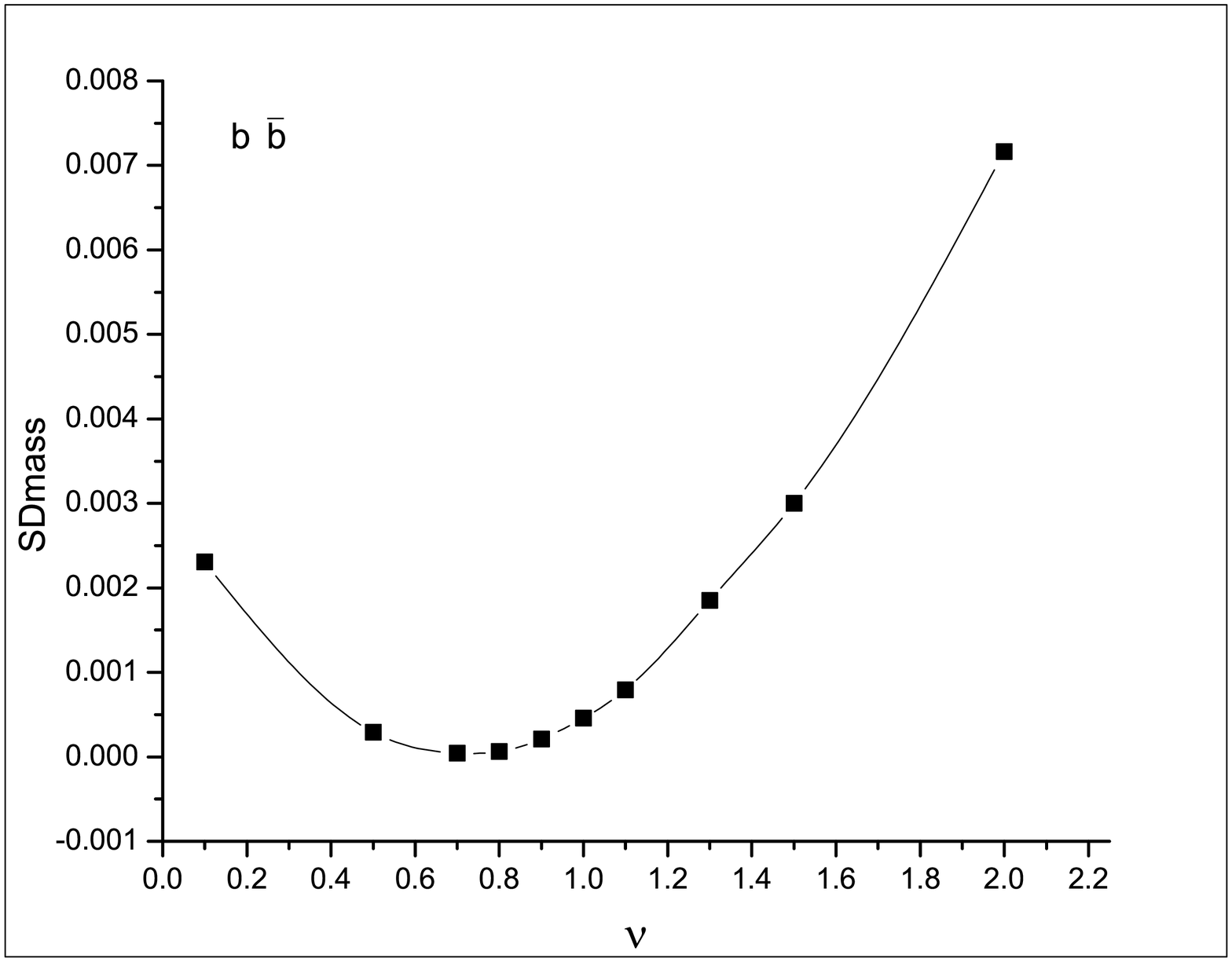}}
\caption{rms deviation in mass for quarkonia states}\label{fig:rmsmass}
\end{center}
\end{figure}

\begin{figure}[htbp]
\begin{center}
\subfigure{\label{fig:photoncc}
\includegraphics[width=6cm, height=4cm, trim=.8cm 1cm 1cm .51cm]{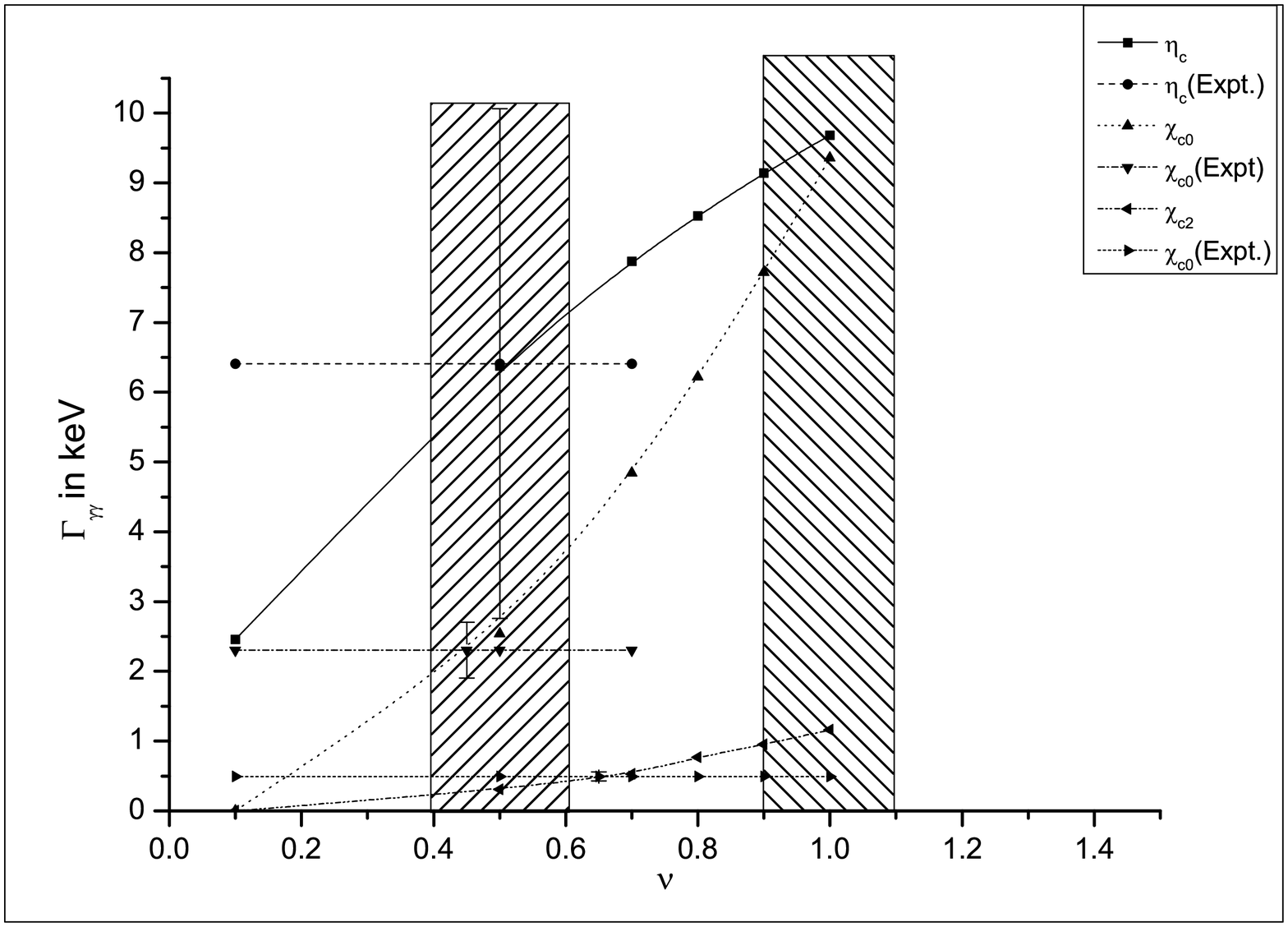}}
\subfigure{\label{fig:gluoncc}
\includegraphics[width=6cm, height=4cm, trim=.8cm 1cm 1cm .51cm]{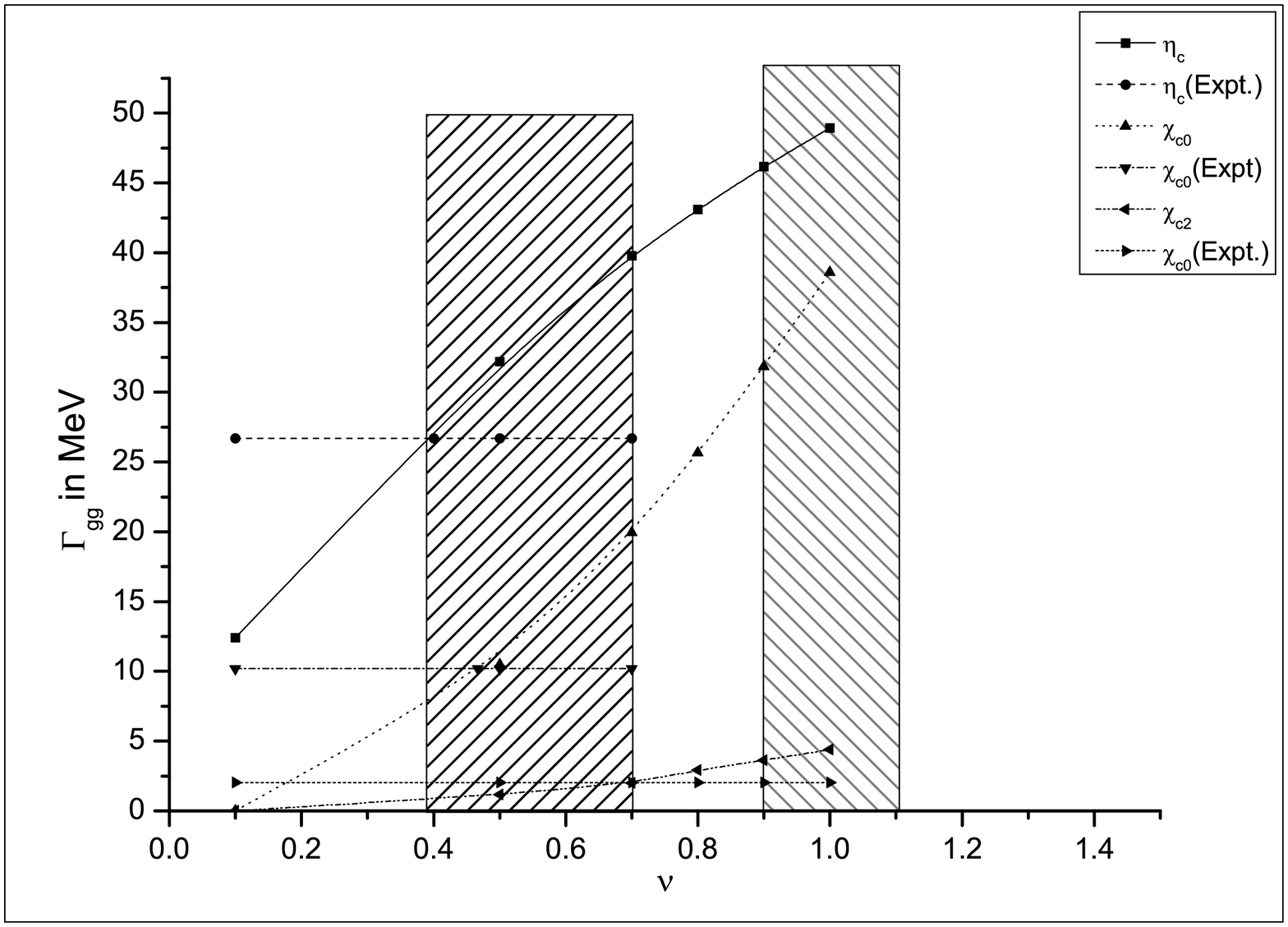}}
\caption{Two-photon and two-gluon decay width of charmonia states against potential index $\nu$}\label{fig:photongluoncc}
\end{center}
\end{figure}

\begin{figure}[!htbp]
\begin{center}
\subfigure{\label{fig:rmsphoton}
\includegraphics[width=5cm, height=4cm, trim=.8cm 1cm 1cm .51cm]{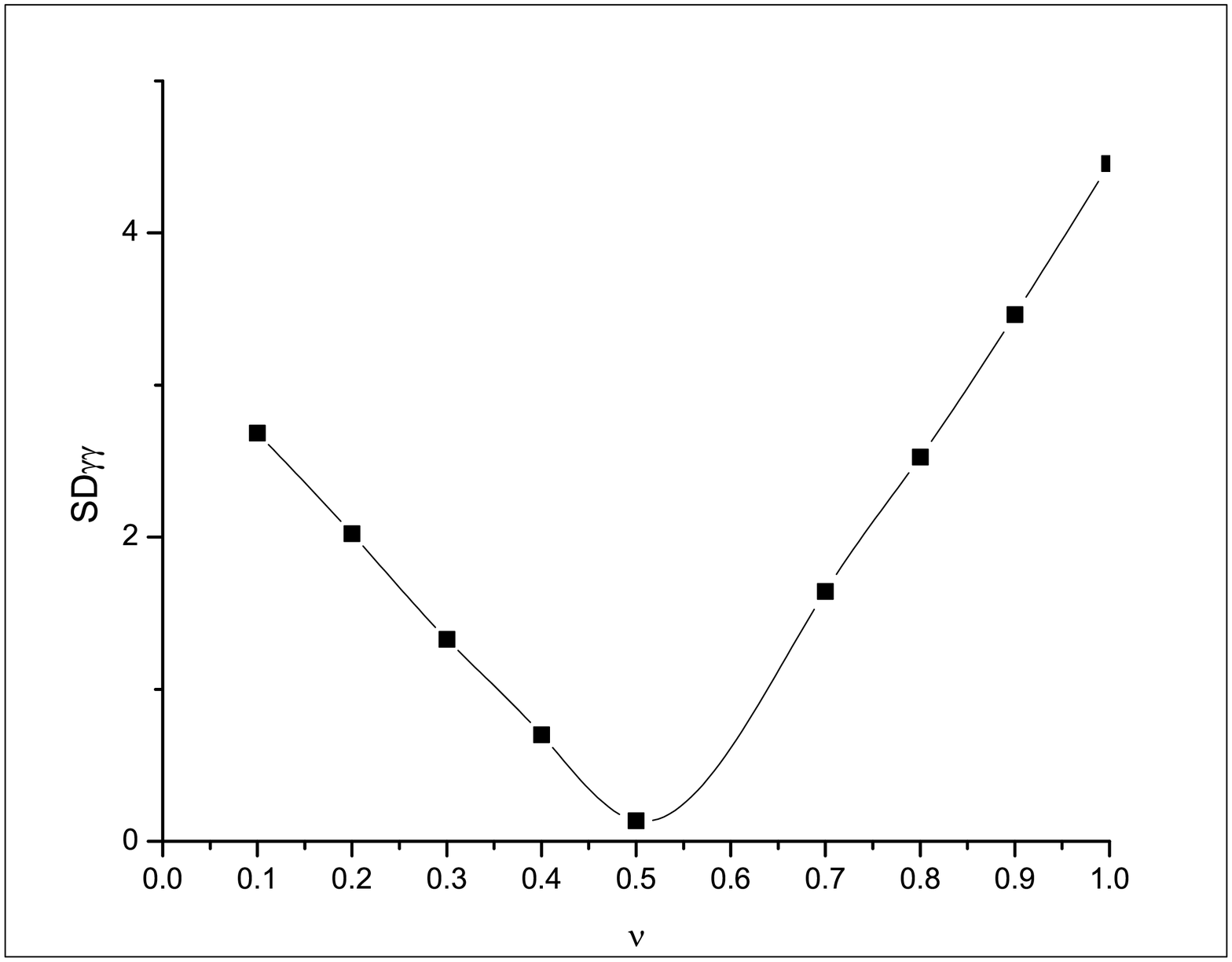}}
\subfigure{\label{fig:rmsgluon}
\includegraphics[width=5cm, height=4cm, trim=.8cm 1cm 1cm .51cm]{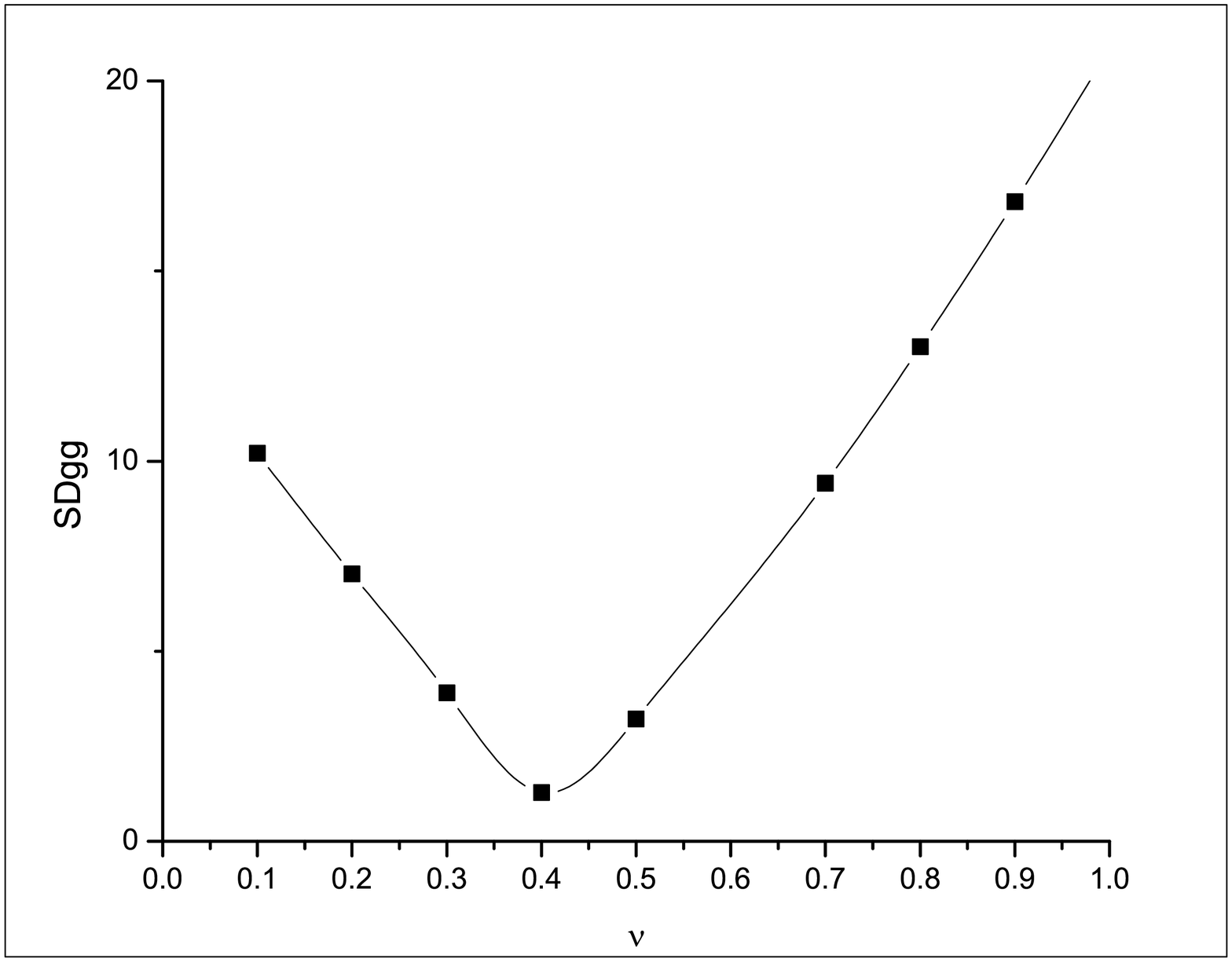}}
\caption{rms deviation in two-photon and two-gluon decay of charmonia}
\end{center}
\end{figure}

\begin{figure}[htbp]
\begin{center}
\subfigure{\label{fig:E1CC}
\includegraphics[width=4cm, height=3cm, trim=.8cm 1cm 1cm .51cm]{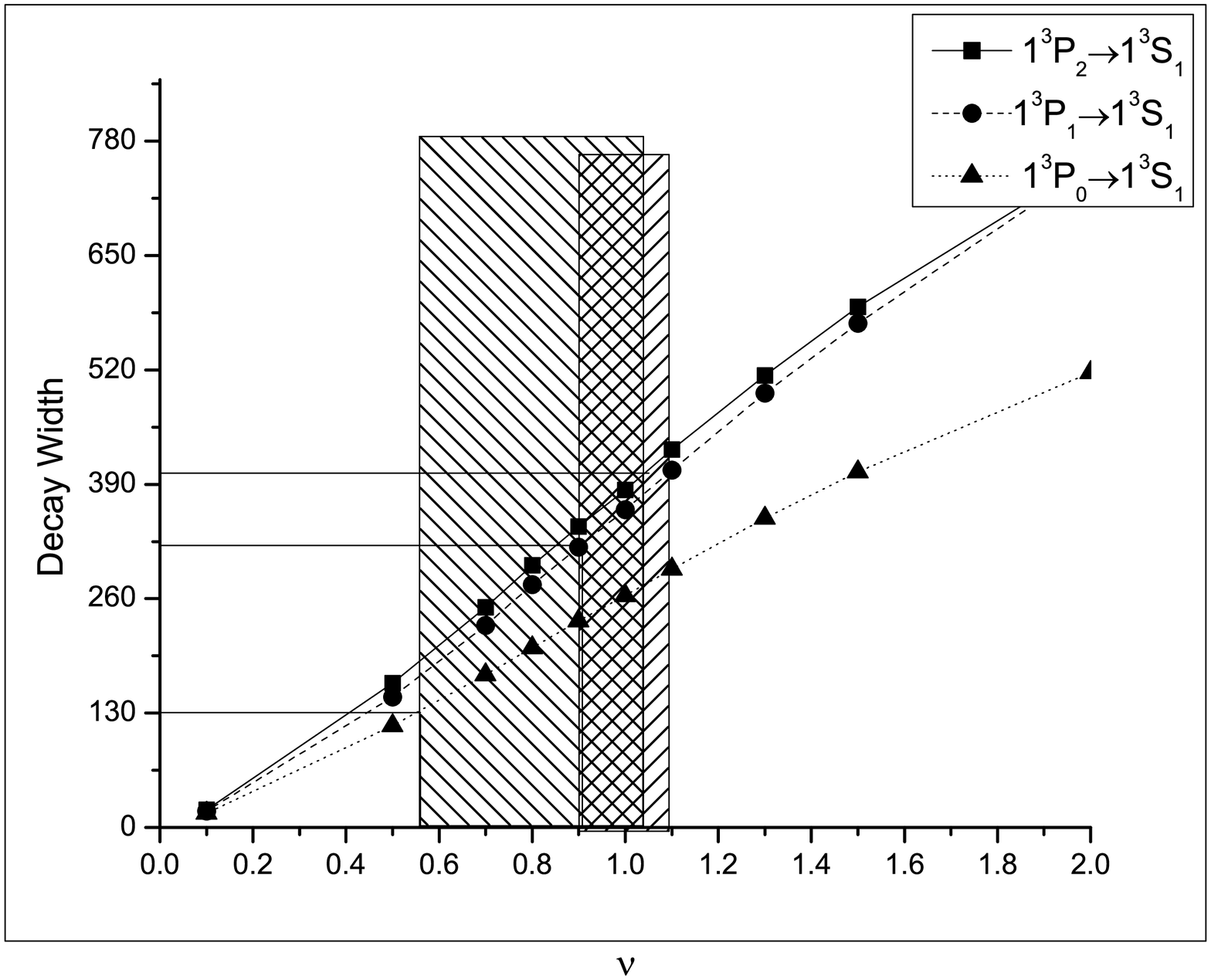}}
\subfigure{\label{fig:E1BC}
\includegraphics[width=4cm, height=3cm, trim=.8cm 1cm 1cm .51cm]{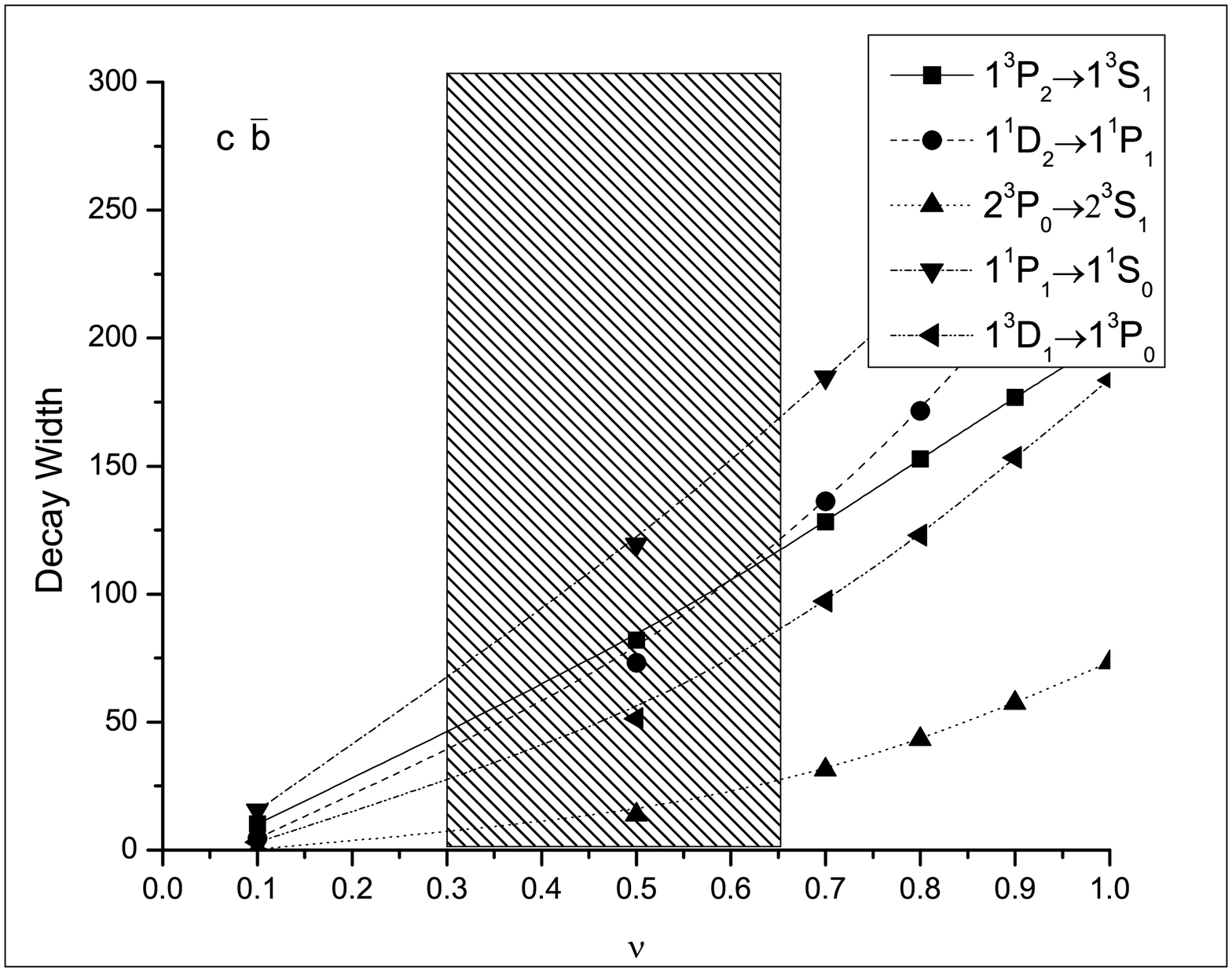}}
\subfigure{\label{fig:E1BB}
\includegraphics[width=4cm, height=3cm, trim=.8cm 1cm 1cm .51cm]{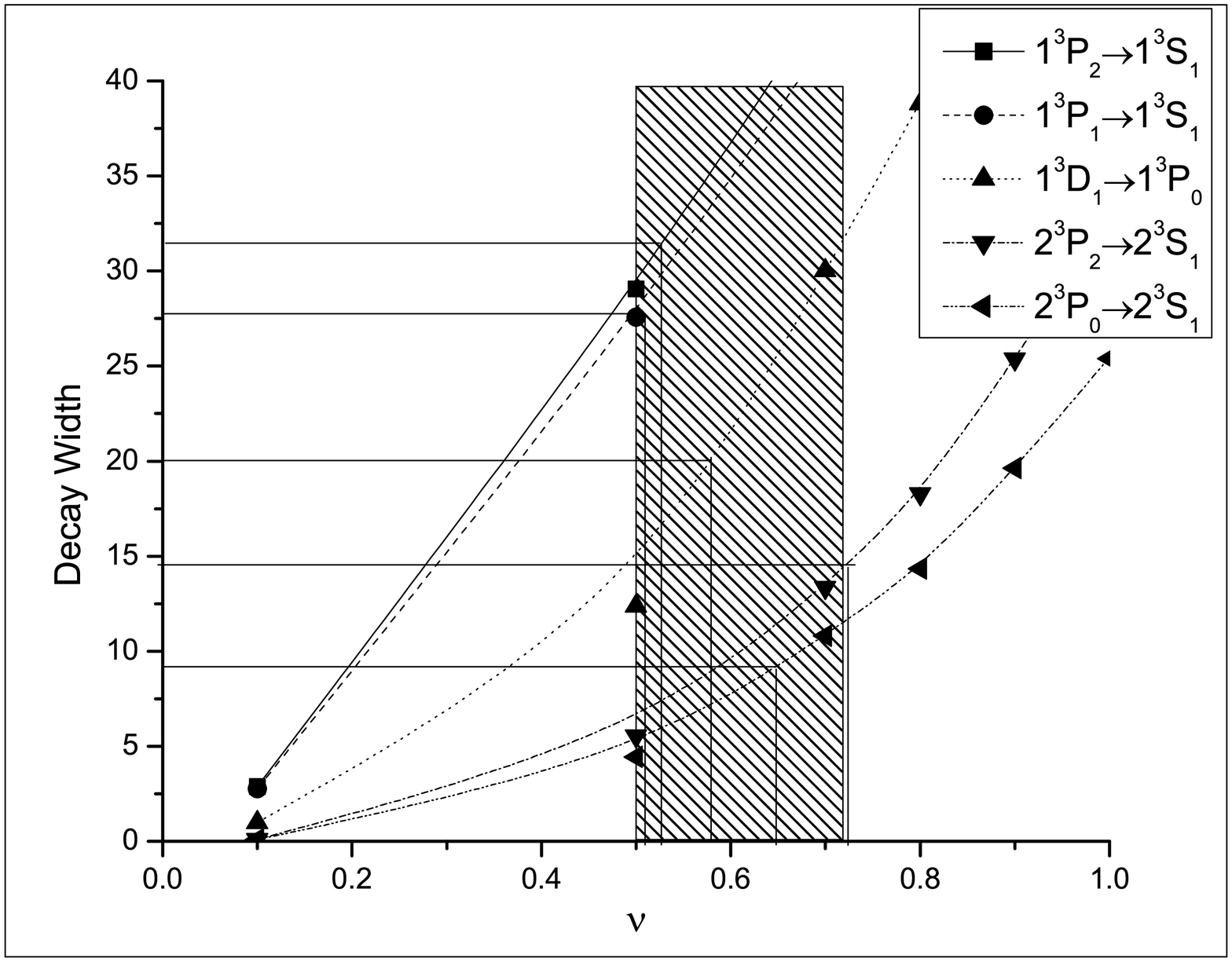}}
\caption{E1 transition decay widths for quarkonium states against potential index $\nu$}\label{fig:E1}
\end{center}
\end{figure}

\begin{figure}[htbp]
\begin{center}
\subfigure{\label{fig:M1CC}
\includegraphics[width=4cm, height=3cm, trim=.8cm 1cm 1cm .51cm]{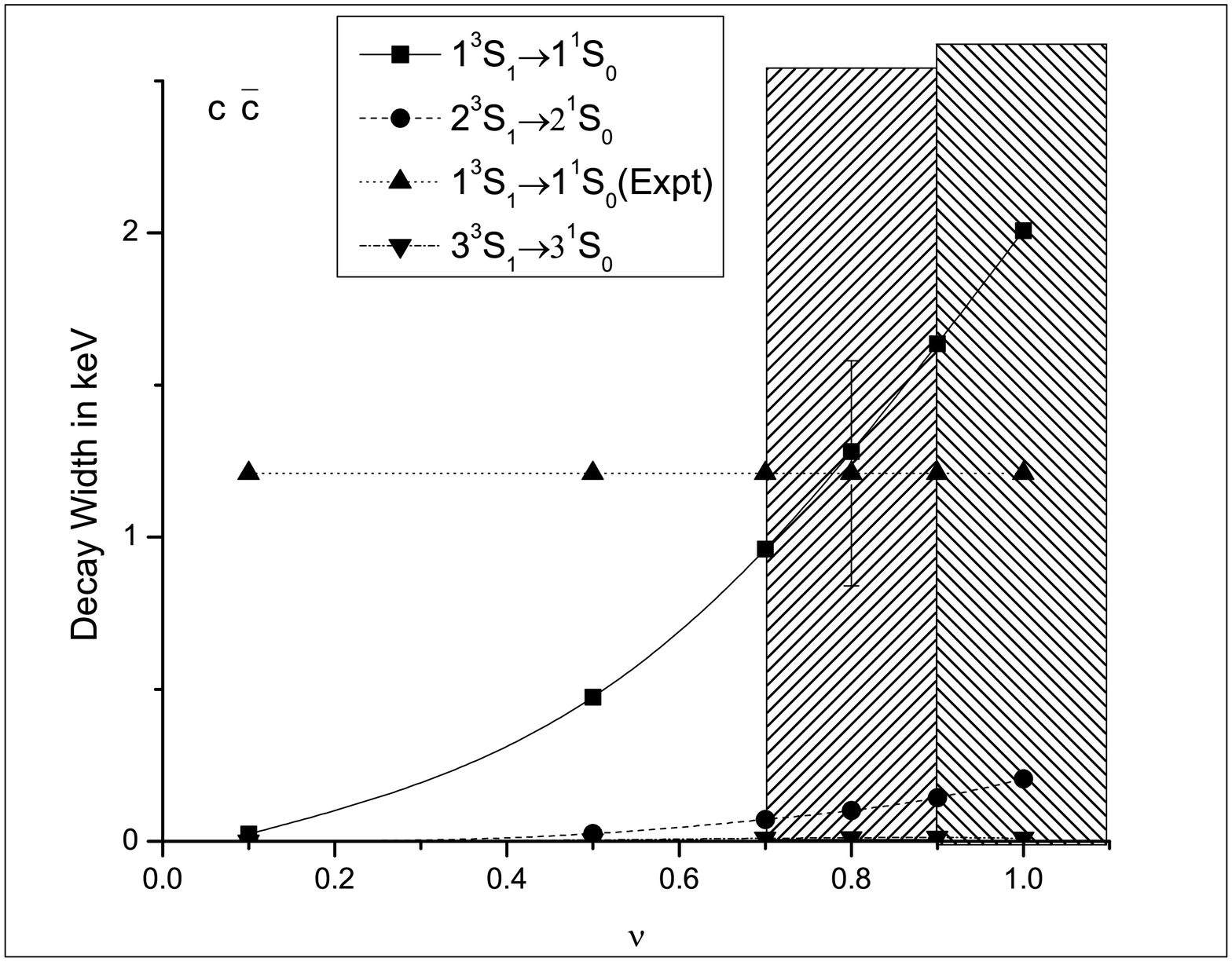}}
\subfigure{\label{fig:M1BC}
\includegraphics[width=4cm, height=3cm, trim=.8cm 1cm 1cm .51cm]{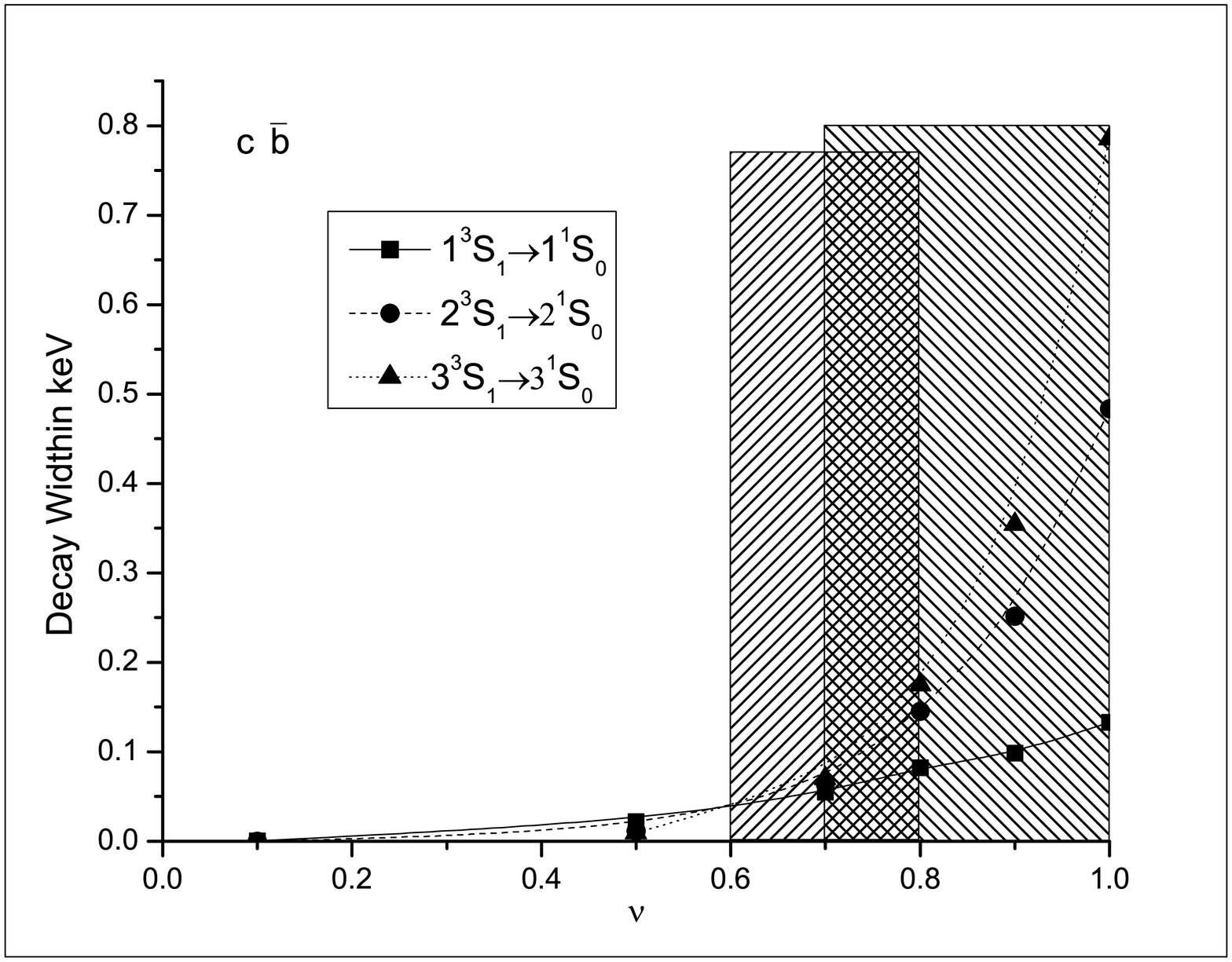}}
\subfigure{\label{fig:M1BB}
\includegraphics[width=4cm, height=3cm, trim=.8cm 1cm 1cm .51cm]{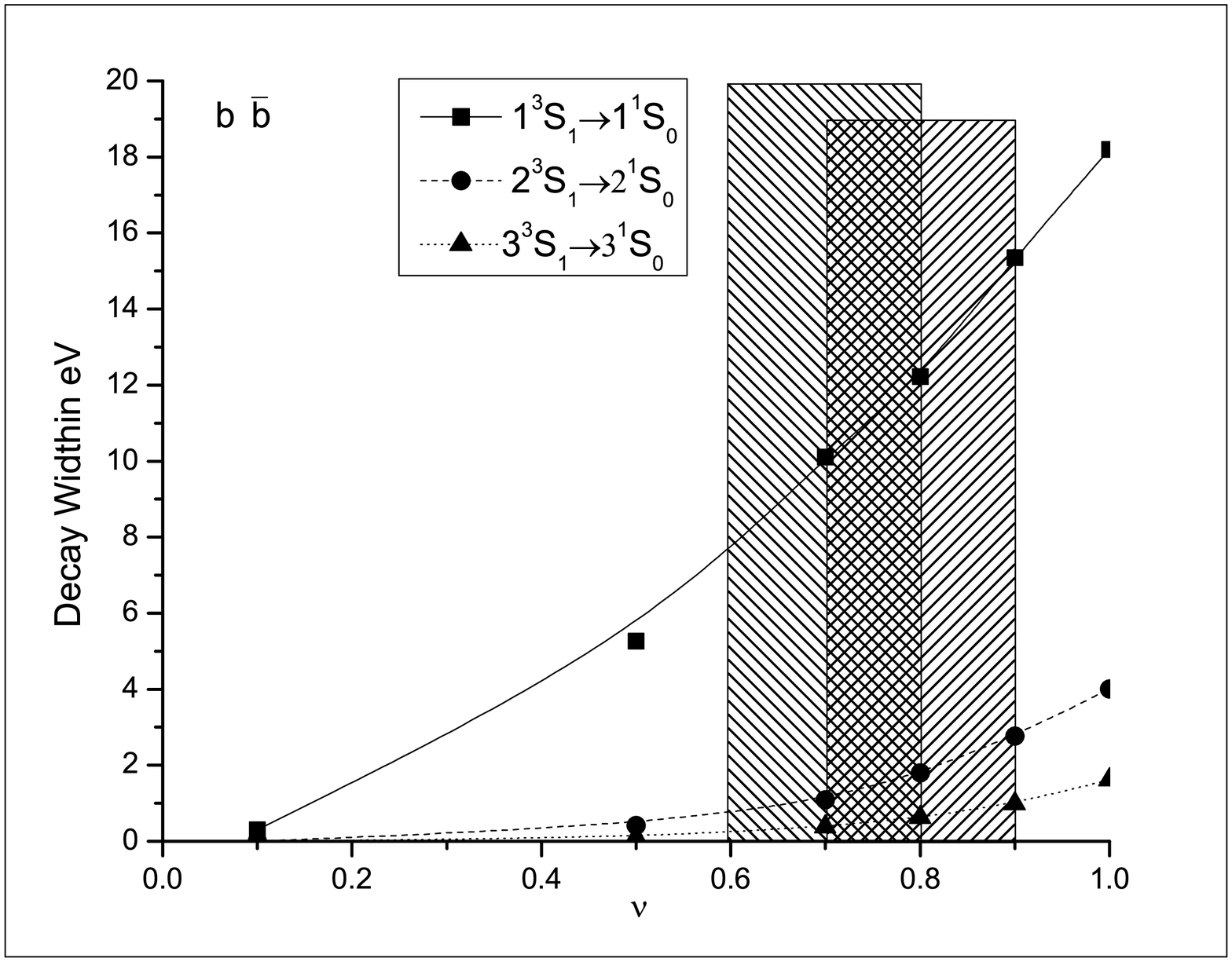}}
\caption{M1 transition decay widths for quarkonium states against potential index $\nu$}\label{fig:M1}
\end{center}
\end{figure}

The computed masses of charmonium, $B_c$ meson and bottonium low lying states using the $CPP_\nu$ model are shown in Table \ref{masscc}, \ref{massbc} and \ref{massbb} respectively with different choices of exponent $\nu$. Their respective experimental values (PDG average) are also listed for comparison. The best possible choice of the exponent ($\nu$) for the description of quarkonia ($b\bar b$ $\&$ $c\bar c$) spectra would be the one with minimum statistical deviations of the predicted mass spectra with the corresponding experimental values. For this we consider about ten experimentally known states ($n^{2S+1}L_J$)  of $c\bar c$ and $b\bar b$ systems as given in Tables \ref{masscc} and \ref{massbb}, and compute the root mean square deviations of the predicted masses of these states for each choices of $\nu$ as
\begin{equation}
SDmass(\nu)=\sqrt{\frac{1}{N}\sum\limits_{i=1}^{N}\left[\frac{M_{CPP_\nu}(n^{2S+1}L_J)-M_{exp}(n^{2S+1}L_J)}{M_{exp}(n^{2S+1}L_J)}\right]_i^2}
\end{equation}
The computed values of the SDmass($\nu$) for $c\bar c$ and $b\bar b$ systems are plotted against $\nu$ in Fig. \ref{fig:rmsmass}. The figure shows distinct minima around $\nu=1.0$ in the case of $c\bar c$ and around $\nu=0.7$ in the case of $b\bar b$ systems. Thus, we conclude that charmonia spectra are better described with the cornell-like potential with $\nu=1.0$, while the bottonia spectra are better described by a relatively flat potential with $\nu=0.7$. In the absence of known experimental excited states for $B_c$ system, we consider that better predictions for the $B_c$ spectra must lie within the range of the exponent $0.7\leq\nu\leq1.0$. And it is found that the predicted masses of the $B_c$ states for the choices of $\nu$ in the range, $0.7\leq\nu\leq0.9$ are in good agreement with other theoretical predictions \cite{Ebert,Vary,Eq7,GLKT}.\\
The computed two-photon and two-gluon decay widths of $\eta_Q$, $\chi_{Q0}$ and $\chi_{Q2}$ states $(Q\in c,b)$ using the spectroscopic parameters of $CPP_\nu$ model are compared with other theoretical predictions in Tables \ref{twophotoncc}, \ref{twophotonbb}, \ref{twogluoncc} and \ref{twogluonbb} respectively. The experimentally known two-photon and two-gluon widths of the $c\bar c$ states are also compared in Table \ref{twophotoncc} and \ref{twogluoncc} respectively. The two-gluon decay process accounts for substantial part of their hadronic decays and hence we have compared our results of the $c\bar c$ system with their measured hadronic decay widths. Though there are many model predictions for the two-photon decay widths of $\chi_{c0}$ and $\chi_{c2}$ states, only very few predictions for the $\eta_c\rightarrow\gamma\gamma$ state exist. Our predictions for $\Gamma_{\gamma\gamma}(c\bar c)$ with the exponent lying in the range, $0.4<\nu<0.7$ and that for $\Gamma_{gg}(c\bar c)$ in the range, $0.3<\nu<0.5$ are in accordance with other model predictions.  Fig. \ref{fig:photongluoncc} shows the trend lines for the two-photon and two-gluon decay widths of the charmonia states with exponent, $\nu$.  The horizontal lines with error bar represents the respective experimental values obtained from the PDG data \cite{PDG2008}. Similar to the rms deviations computed for the mass predictions, the rms deviations of the predicted two-photon widths of the charmonia states with their respective experimental values show a minimum around $\nu=0.5$ (see Fig. \ref{fig:rmsphoton}) while that in the case of two-gluon widths occur around $\nu=0.4$ (See Fig. \ref{fig:rmsgluon}). The shaded regions in Fig. \ref{fig:photongluoncc} with left aligned lines show the neighborhood region of the exponent $\nu$ at which SDmass is minimum for $c\bar c$ states. And the right aligned shaded region show the region of $\nu$ ($0.4<\nu<0.7$) around which
the minimum rms deviations of the two-photon widths ($SD-\gamma\gamma$) occur. The present analysis thus clearly indicates that the mass predictions and the annihilation widths studied here cannot be explained by any single potential choice.\\
\begin{figure}[htbp]
\begin{center}
\subfigure{\label{fig:photonbb}
\includegraphics[width=6cm, height=4cm, trim=.8cm 1cm 1cm .51cm]{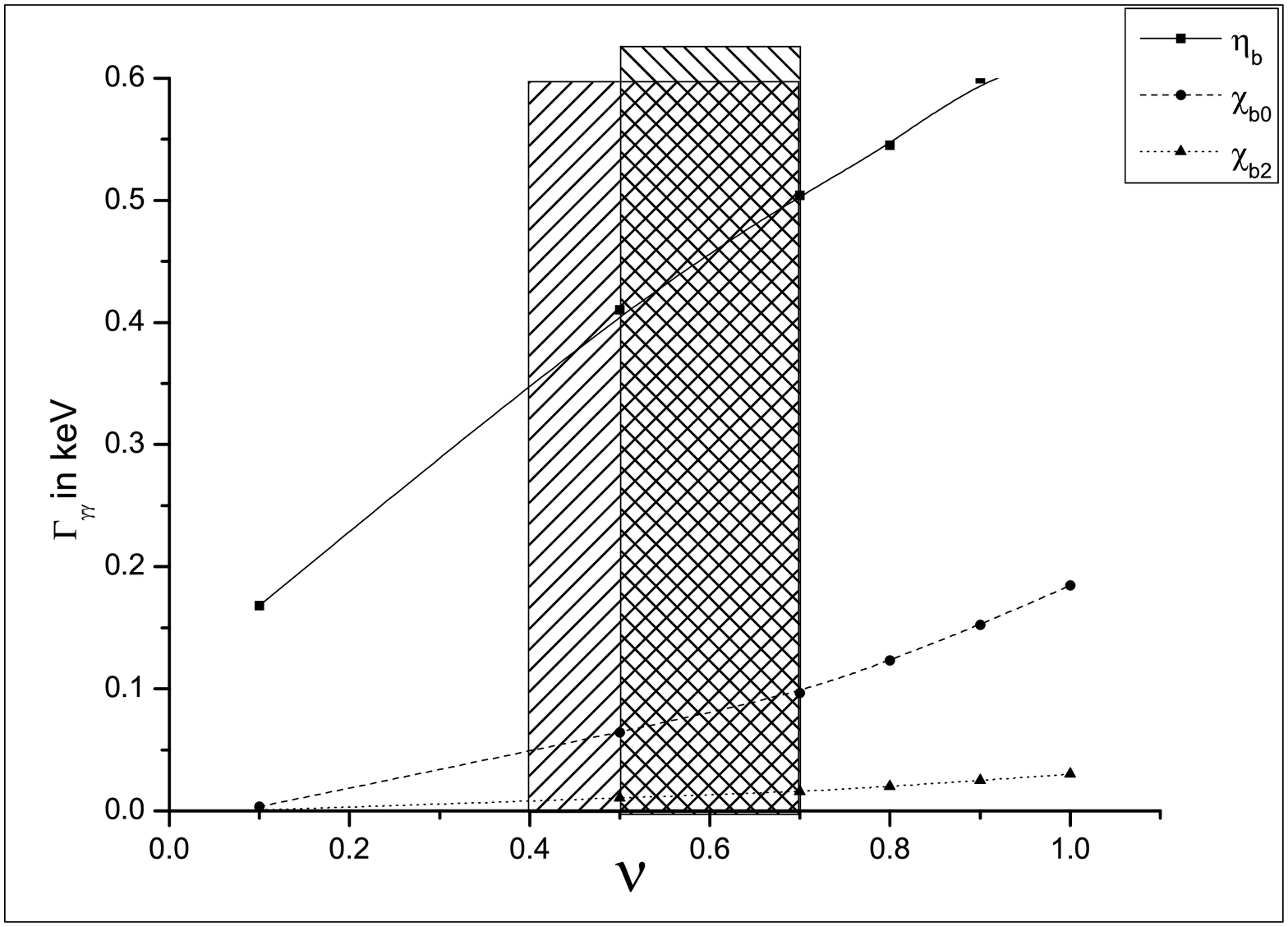}}
\subfigure{\label{fig:gluonbb}
\includegraphics[width=6cm, height=4cm, trim=.8cm 1cm 1cm .51cm]{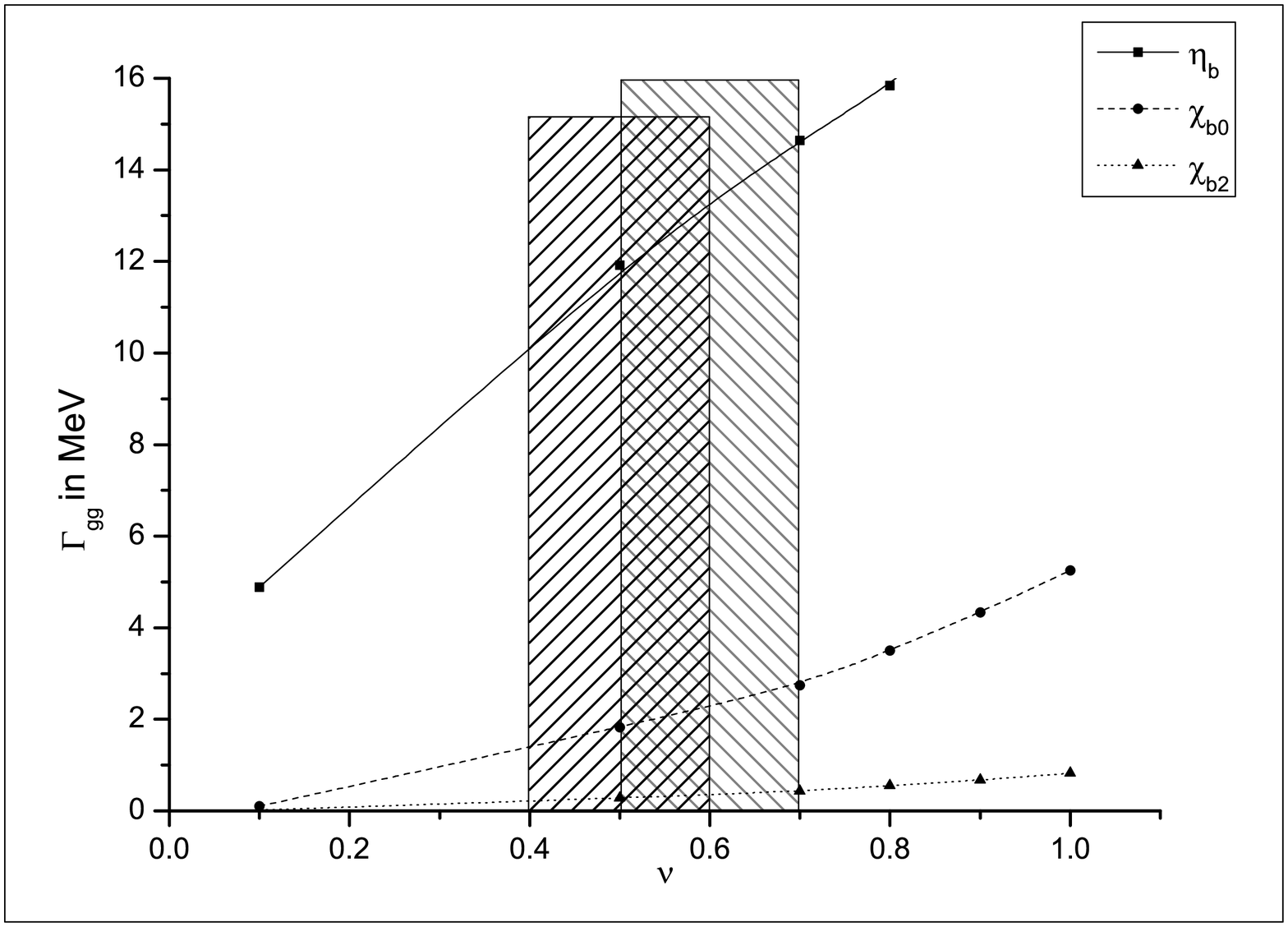}}
\caption{Two-photon and two-gluon decay widths of bottonia states against potential index $\nu$}\label{fig:photongluonbb}
\end{center}
\end{figure}
In the case of $b\bar b$ systems, Fig. \ref{fig:photongluonbb} shows the trend lines for the two-photon and two-gluon decay widths with $\nu$. These decay widths of $\eta_b$, $\chi_{b0}$ and $\chi_{b2}$ states have not been measured. However many other theoretical models predicted the two-photon decay of $\eta_b$ state, while very few model predictions exist for $\chi_{b0}$ and $\chi_{b2}$ states. Our two-photon and two-gluon widths of the $b\bar b$ systems within the exponent, $0.3<\nu\leq0.7$ are in accordance with other theoretical predictions.
New experimental results for these decay widths of $b\bar b$ system are awaited. Present widths in the range of exponent $0.3<\nu<0.7$ for two-photon and two-gluon decays are marked in Fig. \ref{fig:photonbb} as right aligned shaded region.\\
Further, the present study clearly indicates that the $Q\bar Q$ system in its annihilation channel (digamma/digluon decays) interact more weakly compared to their bound state spectral excitations.\\
The low lying electric dipole transition rates of the $c\bar c$, $b\bar b$ and $c\bar b$ systems are listed in Table \ref{E1cc}, \ref{E1bb} and \ref{E1bc} respectively, with other theoretical model predictions and with available experimental results. Their trendlines drawn against $\nu$ are shown in Fig. \ref{fig:E1}. The horizontal lines that cut the E1-curves correspond to the respective experimental values. The shaded region correspond to the neighbourhood of $\nu$ for which the SDmass is minimum. The present results for the E1-transition rates agree with the experimental data \cite{PDG2008} in the neighbourhood region of $\nu=1.0$ in the case of charmonium systems and in the neighbourhood region of $\nu=0.7$ in the case of bottonium systems. These values of $\nu$ are the same at which SDmass becomes minimum.\\
The predicted M1 transition rates with the correction due to the confining exchange current contribution for $Q\bar Q$ states are drawn in Fig. \ref{fig:M1} with respect to the exponent $\nu$. Present result in the case of $J/\psi\rightarrow\eta_c\gamma$ is in agreement with the known experimental result of $1.21\pm0.37$ $keV$ at the exponent $\nu\approx0.8$ with the confining exchange current contribution while it agrees with the prediction at the exponent $\nu\approx0.5$ without the exchange current contribution. Our results for $c\bar c$ system in the range, $0.7\leq\nu\leq1.0$ of the exponent, $\nu$ are in accordance with the values predicted by other models. From the present study, the importance of the two -quark confining exchange current contribution in the M1 transition widths of $c\bar c$ system becomes very clear. In the case of $b\bar b$ and $c\bar b$ states, our predictions with and without the exchange current contribution in the range $0.6<\nu<1.0$ of the exponent, $\nu$ are in accordance with other model predictions. However lack of experimental data for these decay widths does not allow to draw conclusion in favour of a particular choice of the exponent $\nu$. \\
To summarize, we find that the description of the quarkonia mass spectra and the E1 transition can be described by the same interquark model potential of the $CPP_\nu$ with $\nu=1.0$ for $c\bar c$ and  $\nu=0.7$ for $b\bar b$ systems, while the M1 transition (at which the spin of the system changes) and the decay rates in the annihilation channel of quarkonia are better estimated by a shallow potential  with $\nu<1.0$. We look forward to future experimental data related to the decay properties of $c\bar b$ and $b\bar b$ systems.\\
\noindent {\large{\bf Acknowledgment}} \\ Part of this work is
done with a financial support from DST, Government of India,
under a Major Research Project \textbf{SR/S2/HEP-20/2006}.\\\\
\bibliographystyle{model1a-num-names}
\bibliography{<your-bib-database>}

\begin{thebibliography}{99}
\bibitem{Aubert1974}Aubert J J \emph{et al.}, Phys. Rev. Lett. \textbf{33} (1974) 1404.
\bibitem{Augustin1974}Augustin J E \emph{et al.}, Phys. Rev. Lett. \textbf{33} (1974) 1406.
\bibitem{Choi2003}Choi S K \emph{et al.}, (Belle Collaboration), Phys. Rev. Lett. \textbf{91} (2003) 262001.
\bibitem{Ecklund2008}Ecklund K M \emph{et al.}, [CLEO Collaboration], Phys. Rev. \textbf{D 78}, (2008) 091501 arXiv:0803.2869 [hep-ex](2008) and other recent results quoted therein.
\bibitem{PDG2008}Amsler C \emph{et al.}, (Particle Data Group), Phys. Lett. \textbf{B 667} (2008) 1.
\bibitem{Lansberg2006}Lansberg J P, J. Mod. Phys. \textbf{A 21} (2006) 3857 arXiv:hep-ph/0602091.
\bibitem{Lansberg2007}Lansberg J P and Pham T N, Phys. Rev. \textbf{D 75} (2007) 017501.
\bibitem{Brambilla2005}Brambilla N et al, Rev. Mod. Phys. \textbf{77} (2005) 1423, arXiv:hep-ph/0410047.
\bibitem{Eichten5845}Eichten E J and Quigg Chris, Phys. Rev. D \textbf{49} (1994) 5845.
\bibitem{9907240}Brambilla N, Pineda A, Soto J and Vairo A, Nucl. Phys. \textbf{B 566} (2000) 275, hep-ph/9907240.
\bibitem{Barbieri1976}Barbieri R, Gatto R and Kogerler R, Phys. Lett. \textbf{B 60} (1976) 183.
\bibitem{Bodwin1992}Bodwin Geoffrey T., Braaten Eric and Lepage G. Peter, Phys. Rev. \textbf{D 46} (1992) R1914.
\bibitem{Gupta1996}Gupta Suraj N, Johnson James M, Repko Wayne W, Phys. Rev. \textbf{D 54} (1996) 2075, arXiv:hep-ph/9606349.
\bibitem{Huang1996}Huang Han-Wen and Chao Kuang-Ta, Phys. Rev. \textbf{D 54} (1996) 6850;\\ errata Phys. Rev. \textbf{D 56} (1996)1821.
\bibitem{Schuler1998}Schuler G A, Berends F A and Gulik R van, Nucl. Phys. B \textbf{523} (1998) 423 [arXiv:hep-ph/9710462].
\bibitem{Munz1996}Munz C R, Nucl. Phys. A \textbf{609} (1996) 364 [arXiv:hep-ph/9601206].
\bibitem{Crater2006}Crater Horace W, Wong Cheuk-Yin and Alstine Peter Van, Phys. Rev. \textbf{D 74} (2006) 054028.
\bibitem{Wang2007}Wang G-Li, Phys. Lett. \textbf{B 653} (2007) 206.
\bibitem{Laverty2009}Laverty James T, Radford Stanley F and Repko Wayne W, arXiv:0901.3917 [hep-ph].
\bibitem{Godfrey1985}Godfrey Stephen and Isgur Nathan, Phys. Rev. \textbf{D 32} (1985) 189.
\bibitem{Ebert2003}Ebert D, Faustov R N and Galkin V O, Int. Mod. Phys. Lett. \textbf{A 20} (2005) 1887.
\bibitem{Barnes1992}Barnes T, Proceedings of IX InternationalWorkshop on Photon-Photon Collisions, edited by D. O. Caldwell and H.P.Paar(World Scientific, Singapore, 1992) p. 263.
\bibitem{Vairo}Brambilla N, Mereghetti E and Vairo A, Phys. Rev. \textbf{D 79} (2009) 074002 arXiv:0810.2259.
\bibitem{AKRai2005}Rai A K, Pandya J N and P C Vinodkumar, J. Phys. G : Nucl. Part. Phys. \textbf{31} (2005) 1453.
\bibitem{AKRai2008}Rai A K, Pandya J N and P C Vinodkumar, Eur. Phys. J. \textbf{A 38} (2008) 77 arXiv:0901.1546[hep-ph].
\bibitem{Ebert}Ebert D, Faustov R N and Galkin V O, Phys. Rev. \textbf{D 67} (2003) 014027.
\bibitem{Ebert200318}Ebert D, Faustov R N, and Galkin V O, Mod. Phys. Lett. \textbf{A 18} (2003) 601.
\bibitem{Kim2005}Kim C S, Lee T, and Wang G L, Phys. Lett. \textbf{B 606}, (2005) 323.
\bibitem{rungekutta}Lucha W and Shoberl F, Int. J. Mod. Phys. \textbf{ C 10} (1999), arXiv:hep-ph[9811453].
\bibitem{Branes2005}Branes T, Godfrey S and Swanson E S, Phys.Rev. \textbf{D72} (2005) 054026.
\bibitem{Lakhina2006}Lakhina Olga and Swanson Eric S, Phys. Rev \textbf{D 74} (2006) 014012 [arXiv:hep-ph/0603164].
\bibitem{Voloshin2008}Voloshin M B, Prog. Part. Nucl. Phys. \textbf{61} (2008) 455 arXiv:hep-ph/0711.4556.
\bibitem{Eichten2008}Eichten E, Godfrey S, Mahlke H and Rosner J L, Rev. Mod. Phys. \textbf{80} (2008) 1161.
\bibitem{Gerstein1995}Gershtein S S, Kiselev V V, Likhoded A K and Tkabladze A V, Phys. Rev. {\bf D51} (1995) 3613.
\bibitem{Landau1949}Landau L, Phys. Abstracts \textbf{A 52} (1949) 125.
\bibitem{Yang1950}Yang C N, Phys. Rev. \textbf{77} (1950) 242.
\bibitem{Zou2003}Zou B S and Chiang H C, Phys. Rev. \textbf{D 69} (2003) 034004.
\bibitem{Bhavin2009}Patel Bhavin \emph{et al.}, J. Phys. G: Nucl. Part. Phys. \textbf{36} (2009) 035003.
\bibitem{Petrelli1998}Petrelli A, Cacciari M, Greco M, Maltoni F, Mangano M L, Nucl. Phys. \textbf{B 514} (1998) 245.
\bibitem{Kwong1988}Kwong Waikwok et. al., Phys. Rev. \textbf{D 37} (1988) 3210.
\bibitem{Vary}Abd El-Hady A, Spence J R and Vary J P, Phys. Rev. D \textbf{71} (2005) 034006.
\bibitem{Eq7}Eichten E and Quigg C, Phys. Rev. D \textbf{49} (1994) 5845.
\bibitem{GLKT}Gershtein S S, Kiselev V V, Likhoded A K and Tkabladze A V, Phys. Usp. \textbf{38} (1995) 1.
\bibitem{Godfrey}Godfrey Stephen, Phys. Rev. D \textbf{70} (2005) 054017.
\bibitem{Barbieri1981}Barbieri R, Caffo M, Gatto R, and Remiddi E, Nucl. Phys. \textbf{B 192} (1981) 61.
\bibitem{Mangano1995}Mangano M and Petrelli A, Phys. Lett. \textbf{B 352} (1995) 445.
\bibitem{Lansberg2009}Lansberg J P and Pham T N, Phys. Rev. \textbf{D 79} (2009) 094016, arXiv:0903.1562[hep-ph].
\bibitem{Eichten3090}Eichten E \emph{et al.}, Phys. Rev. D \textbf{17} (1978) 3090.
\bibitem{Floriana}Giannuzzi Floriana, Phys. Rev. \textbf{D 78} (2008) 117501.
\bibitem{Fabiano2003}Fabiano N, Eur. Phys. J.  \textbf{C 26} (2003) 441 [arXiv:hep-ph/0209283].
\bibitem{Ackleh1992}Ackleh E S and Barnes T, Phys. Rev. D \textbf{45} (1992) 232.
\bibitem{Ahmady1995}Ahmady Mohammad R and Mendel Roberto R, Phys. Rev. D \textbf{51} (1995) 141 [arXiv:hep-ph/9401315].
\bibitem{lansbergpham}Lansberg J P and Pham T N, arXiv:0804.2180[hep-ph].
\bibitem{Radford}Radford S F and Repko W W, Phys. Rev. D \textbf{75} (2007) 074031.
\bibitem{0412158}Heavy Quarkonium Physics, N Brambill \emph{et al.}, CERN Yellow Report, CERN 2005-005, Geneva: CERN, arxiv:hep-ph/0412158.
\bibitem{Lahde}Lahde T A, Nucl. Phys. A \textbf{714} (2003) 183.
%\bibitem{Ebert}Ebert D, Faustov R N and Galkin V O, Phys. Rev. D \textbf{67} (2003) 014027.
\bibitem{Lahde645}Lahde T A \emph{et al.}, Nucl. Phys. \textbf{A} \textbf{645} (1999) 587.
\bibitem{PRD.30.1924}Grotch H, Owen D A and Sebastian K J, Phys. Rev. \textbf{D 30}, (1984), 1924.
\bibitem{Gershtein3613}Gershtein S S, Kiselev V V, Likhoded A K and Tkabladze A V, Phys. Rev. D \textbf{51} (1995) 3613.








\end{thebibliography}

\end{document}